\documentclass[prx,onecolumn,nofootinbib,citeautoscript,eqsecnum,10pt,notitlepage]{revtex4-2}
\pdfoutput=1

\usepackage{dsfont}
\usepackage{float}
\usepackage{array}
\usepackage{slashed,bbold}
\usepackage{physics}
\usepackage{enumerate}

\usepackage{amsmath,amssymb,bm} 
\usepackage{graphicx}

\usepackage[tight]{subfigure} 

\usepackage{color} 
\usepackage[papersize={8.5in,11in}]{geometry}

\usepackage{color}
\definecolor{darkblue}{rgb}{0.,0.,0.4}
\definecolor{darkred}{rgb}{0.5,0.,0.}
\definecolor{BlueViolet}{RGB}{138,43,226}
\definecolor{SkyBlue}{RGB}{30,144,255}
\definecolor{DarkGreen}{RGB}{0,100,0}
\usepackage[pdftex,colorlinks=true,linkcolor=darkblue,citecolor=blue,urlcolor=darkred]{hyperref}

\def \be{\begin{equation}}
\def \ee{\end{equation}}
\def \nn{\nonumber \\}
\begin{document}

\title{Transmission and conductance across junctions of isotropic and anisotropic three-dimensional semimetals}

\author{Ipsita Mandal}
\affiliation{Department of Physics, Shiv Nadar Institution of Eminence (SNIoE), Gautam Buddha Nagar, Uttar Pradesh 201314, India}

\begin{abstract}
We compute the transmission coefficients and zero-temperature conductance for chiral quasiparticles propagating through various geometries, which consist of junctions of three-dimensional nodal-point semimetals. In the first scenario, we consider a potential step with two Rarita-Schwinger-Weyl or two birefringent semimetals, which are tilted with respect to the other on the two sides of the junction. The second set-up consists of a junction between a doped Dirac semimetal and a ferromagnetic Weyl semimetal, where an intrinsic magnetization present in the latter splits the doubly-degenerate Dirac node into a pair of Weyl nodes. A scalar potential is also applied in the region where the Weyl semimetal phase exists. Finally, we study sandwiches of Weyl/multi-Weyl semimetals, with the middle region being subjected to both scalar and vector potentials. Our results show that a nonzero transmission spectrum exists where the areas, enclosed by the Fermi surface projections (in the plane perpendicular to the propagation axis) of the incidence and transmission regions, overlap.
Such features can help engineer unidirectional carrier propagation, topologically protected against impurity backscattering, because of the chiral nature of the charge carriers.
\end{abstract}

\maketitle

\tableofcontents

%%%%%%%%%%%%%%%%%%%%%%%%%%%%%%%%%%%%%%%%%%%%%%%%%

\section{Introduction}

A wide variety of gapless topological phases have been discovered in recent years, that arise when the Brillouin zone (BZ) harbours two or more bands crossing at a point \cite{bernevig,bernevig2}, and the nontrivial topology in the momentum space results in bandstructures exhibiting nonzero Chern numbers. These are called semimetals due to the existence of the gapless nodal points where the density of states goes to zero. Some of these have high-energy counterparts (e.g. Weyl semimetals), and some do not (e.g. double-Weyl and triple-Weyl semimetals). 
In particular, for multifold semimetals with isotropic linear dispersion, the low-energy effective Hamiltonian can be expressed as $\sim \mathbf{k} \cdot \boldsymbol{\mathcal{S}}$,
where $\boldsymbol{\mathcal{S}}$ represents the vector consisting of the matrices for a particular value of pseudospin --- which are thus natural generalizations of the Weyl semimetal Hamiltonian $\sim \mathbf{k} \cdot \boldsymbol{\sigma}$.\footnote{As usual, $\boldsymbol{\sigma}$ represents the vector of the three Pauli matrices, such that the Weyl semimetal hosts pseudospin-1/2 quasiparticles.} Examples of pseudospin-3/2 quasiparticles include Rarita-Schwinger-Weyl (RSW) semimetals \cite{long} and birefringent semimetals \cite{malcolm,malcolm-bitan,ips-biref}.
On the other hand, multi-Weyl semimetals are anisotropic generalizations of the Weyl semimetals, where the dispersion is linear in one direction (let us call it $k_z$), and proportional to $k_\perp^J$ (with $J=2$ or $J=3$) in the plane perpendicular to it (where $k_\perp =\sqrt{k_x^2+k_y^2}$).\footnote{We note that $J=1$ for Weyl semimetals, making the dispersion isotropic.}

In the branch of high-energy physics, the Rarita-Schwinger (RS) equation describes the field equation for relativistic spin-3/2 fermions. However, no elementary particle with spin-3/2 has been found experimentally and, although they are postulated to exist in models based on supergravity \cite{weinberg}, they do not appear in the standard model. On the other hand, an analogue of these relativistic spin-3/2 fermions exists in the form of quasiparticles carrying pseudospin-3/2 in condensed matter systems, which of course are non-relativistic~\cite{long,igor,igor2,ips3by2}. In an effective Hamiltonian of the form shown in Eq.~\eqref{hamrsw0} (that is discussed in Sec.~\ref{secrswbdirac}), the four bands show linear-in-momentum dispersions fixed to the values $\pm 3 \,|\mathbf k|/2 $ and $\pm |\mathbf k|/2 $ [cf. Eq.~\eqref{eqdisrsw}].
%%%%%
The so-called birefringent semimetals also serve as non-relativistic pseudospin analogues of the elusive relativisic RS fermions. But, unlike the RSW case, the slopes of the linear-in-momentum dispersions of the four bands [cf. Eqs.~\eqref{hambiref1} and \eqref{eqevbiref}] can be continuously tuned by a parameter (labelled as $\zeta $ in this paper).
% PHYSICAL REVIEW B 99, 245126 (2019)

The anisotropic dispersion in a node of a multi-Weyl semimetal (i.e., with $J>1$) results due to the merging of two or more Weyl nodes with the same chirality \cite{bernevig2}, the merger being topologically protected by point-group symmetries (such as C$_4$ and $C_6$ rotational symmetries). Therefore, although the existence of a Weyl semimetal (with $J=1$) does not require
any symmetry (except for the lattice translation symmetry of a crystal), the multi-Weyl semimetals are stabilized with the help of crystalline symmetries \cite{bernevig2,PhysRevLett.107.186806}.

The semimetals described above have been predicted to exist in various candidate materials, based on first principles and density functional theory calculations. For example, (a) Weyl semimetals have been observed in the TaAs family \cite{Huang_2015,Xu_2015} and SrSi$_2$ \cite{Huang1180}; (b) double-Weyl quasiparticles have been predicted to be hosted in HgCr$_2$Se$_4$ \cite{bernevig2,PhysRevLett.107.186806}, SrSi$_2$ \cite{Huang1180}, and superconducting states of $^3$He-A \cite{volovik}, UPt$_3$ \cite{PhysRevB.92.214504}, SrPtAs \cite{PhysRevB.89.020509}, and YPtBi \cite{PhysRevB.99.054505}; (c) molybdenum monochalcogenide compounds A(MoX)$_3$ (where A $ =$Na, K, Rb, In, Tl; X = S, Se, Te) \cite{PhysRevX.7.021019} are expected to harbour triple-Weyl quasiparticles.
%%%%%%%%%%%%%%%
Similarly, the signatures of the RSW semimetals are believed to be associated with the large topological charges found in various materials like CoSi~\cite{takane}, RhSi~\cite{sanchez}, AlPt~\cite{schroter}, and PdBiSe~\cite{ding}. Last but not the least, various systems having strong spin-orbit couplings
\cite{bradlyn,prb.94.195205} (e.g., the antiperovskite family \cite{PhysRevB.90.081112} with the chemical formula A$_3$BX, and the CaAgBi-family materials with a stuffed Wurtzite structure \cite{prm.1.044201})
can host birefringent semimetals.

\begin{figure}[t]
\includegraphics[width = 0.4 \textwidth]{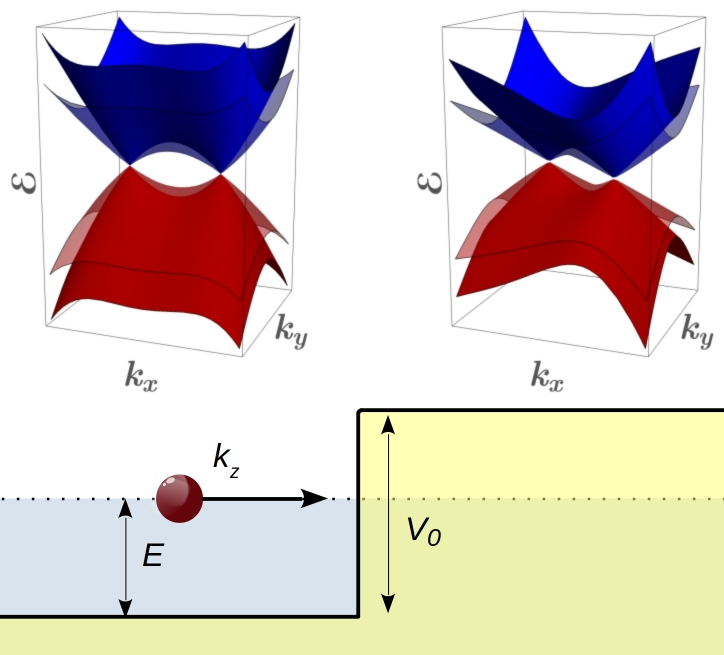}
\caption{\label{figsetup1}
Schematic diagram of the transport of a quasiparticle (red ball) of a four-band isotropic semimetal into a potential step of strength $V_0$. The Fermi level (at energy $E$) is depicted by a dotted line, which cuts the conduction and valence bands on the left and right of the junction, respectively.
Inside the potential step, the cones are rotated in the $xy$-plane by a tilt angle $\theta$.
The energy dispersions (indicated by $\mathcal E$) of the mutually rotated cones, plotted against the
$k_xk_y$-plane, are also shown in the top panel.
}
\end{figure}

With the increase in sophistication of experiments, it is possible to engineer complicated geometries with various semimetal components. This necessitates the theoretical explorations for the signatures that can be found from such set-ups. In this paper, we study the behavior of the transmission coefficients and conductivity across junctions of semimetals. In particular, we consider the following set-ups:
%%%%%%%%%%%
\begin{enumerate}[{(1)}]

\item A junction built with two identical four-band semimetals (RSW or birefringent), but which are tilted with respect to each other across a potential step defining the interface (similar to the set-up considered in Ref.~\cite{tilted_jn} with Weyl semimetals). This is shown in Fig.~\ref{figsetup1}. The motivation for studying this set-up is as follows. In a generic situation, an interface may be created in a semimetallic sample when defects are present, for instance, in the form of irregular surfaces or misaligned lattice structures extending in adjacent regions. Certain experiments may involve a sudden cooling of a sample, which may generate tiny cracks in the material, again causing a misaligned lattice structure on the two sides of such a crack. In all these situations, the lattice translation invariance in the direction perpendicular to the interface is broken, which will case an electron beam travelling across the interface to undergo refraction \cite{Yang_2019}.

\item A junction between a doped Dirac semimetal and a ferromagnetic Weyl semimetal, where an instrinsic magnetization present in the latter splits the degenerate Dirac nodes into a pair of momentum-separated Weyl nodes (cf. Ref.~\cite{dirac-weyl}). A scalar potential is also applied in the region where the Weyl semimetal phase exists. This is shown in Fig.~\ref{figsetup2}.
This set-up is motivated by the fact that a sharp domain wall may exist between two regions with different values of the  intrinsic magnetization.

\item Sandwiches of Weyl/multi-Weyl semimetals consisting of three different $J$-values, with the middle region
being subjected to both scalar and vector potentials (generalization of the set-ups considered in Refs.~\cite{Zhu2020,Deng2020,krish-sandwich,ips-aritra}). This is shown in Fig.~\ref{figsetup3}.
%%%%%%%5
In real experiments, it is possible to generate the vector potential in the potential barrier region
\cite{PhysRevLett.72.1518,Zhai_2008,Ramezani_Masir_2009} by placing a ferromagnetic metal strip deposited on the top of a thin dielectric layer, positioned above the middle slab (of semimetal). The magnetization of the strip should be along the propagation direction of the current, such that the resulting fringe fields provide a magnetic field modulation along the current, which is assumed to be homogeneous in the perpendicular directions.
The motivation for studying these junctions consisting of different $J$-values on the two sides is that this allows the monopole charge (equal to the Chern number) to change across the junction, since a node of these semimetals harbours a monopole charge of magnitude $J$.
In fact, it is a well-known fact that transport measurements across junctions of materials provide access to the topologically nontrivial features of their bandstructures in the Brillouin zone. This can be easily seen from the fact the transmission/reflection coefficients and the conductivity expressions depend on the values of the monopole charges of the semimetals considered \cite{krish-sandwich,Deng2020,ips-aritra,ips-sandip-sajid}.

\end{enumerate}
%%%%%%%%%%%%%%%%%%%%%%%%%
Different semimetals have been chose to represent semimetals of different pseudospin values. Moreover, higher psueudospin values involve higher number of bands describing the semimetal and, also, higher values of monopole charge \cite{igor,ips-cd} at a given nodal point (where the bands cross). Hence, there are representatives of various kinds of semimetals that we have in nature. For Sec.~\ref{secrswbdirac}, since the case of Weyl semimetals (representative of pseudospin-1/2 quasiparticles arising in two-band models) has already been considered in Ref.~\cite{tilted_jn}, we have focussed on the four-band pseudospin-3/2 cases here.
Sec.~\ref{secdiracweyl} discusses the set-up possible with a semimetal which has an intrinsic magnetization, and hence an appropriate example has been considered there. Finally, in Sec.~\ref{secmultiweyl}, we have focussed on sandwiches made of junctions consisting of systems which are all Weyl-like, consisting of two bands, and harbouring quasiparticles of pseudospin-1/2. One could also have considered such sandwiches with two different four-band semimetals on the two sides of a junction, so that both regions have pseudospin-3/2 quasiparticles as the charge carriers, but the computations would have been very cumbersome in those situations. Hence, we have chosen to consider two-band Weyl-like cases, which suffice to highlight some generic characteristics.

In our computations, we use the usual Landauer-B\"{u}ttiker formalism (see, for example, Refs.~\cite{salehi,beenakker,ips-qbcp-tunnel,ips-tunnel-linear,ips-aritra}). We determine the transmission coefficients for the propagation of the quasiparticles in a slab of square cross-section, with a transverse width $W$, where $W$ is assumed to be large enough to impose periodic boundary conditions along these transverse directions. For all the scenarios considered in this paper, the propagation direction is along the $z$-direction, with the quasiparticle incident from the leftmost region of the junction configuration (effectively, coming from $z=-\infty$), and propagating towards the the right (i.e., along the positive $z$-axis.)

The paper is organized as follows. In Sec.~\ref{secrswbdirac}, we consider the geometrical set-ups in
items (1) and (2) described above. Sec.~\ref{secdiracweyl} is devoted to item (3). The sandwich configurations of Weyl/multi-Weyl semimetals are studied in Sec.~\ref{secmultiweyl} [cf. item (4)]. Finally, we end with a summary and outlook in Sec.~\ref{secsum}. In the Appendix, we review the derivation of the boundary conditions at a junction, and show how the continuity of the probability flux dictates the forms of the transmission and reflection coefficients.

%%%%%%%%%%%%%%%%%%%%%%%%%%%%%%%%%%%%%%%%%%%%%%%
%%%%%%%%%%%%%%%%

%%%%%%%%%%%%%%%%%%%%%%%%%%%%%%%%%%%%%
\section{Junction with mutually tilted semimetals}
\label{secrswbdirac}

In this section, we consider a potential step built with the same semimetal (with linear isotropic dispersions near the nodal-points) on both sides, but the cones are tilted with respect to each other across the junction. The set-up is schematically shown in Fig.~\ref{figsetup1}. In particular, we focus on the RSW and birefringent semimetals, both of which feature four bands crossing at each nodal-point.

%%%%%%%%%%%%%%%%%%%%%%%%
\subsection{RSW semimetal}

Crystal structures belonging to the eight space groups $207$-$214$ can host fourfold topological degeneracies about the $\Gamma $, R, and/or H points \cite{bernevig}, and can lead to the linearized $\mathbf{k} \cdot \mathbf {p}$ Hamiltonian (about such a point) of the form
\begin{align}
\label{hamrsw0}
\mathcal{H}_{ \text{RSW}}^{1\text{node}}(\mathbf  k) 
= v_g \,\mathbf{k}\cdot \mathbf J\,,
\end{align}
where $v_g $ is the isotropic Fermi velocity. 
The system hosts pseudospin-3/2 RSW quasiparticles because
the three components of $\mathbf J$ form the spin-3/2 representation of the SO(3) group. A standard representation of $\mathbf J$ is given by
\begin{align}
\label{eqrswxJ}
J_x = \left(
\begin{array}{cccc}
 0 & \frac{\sqrt{3}}{2} & 0 & 0 \\
 \frac{\sqrt{3}}{2} & 0 & 1 & 0 \\
 0 & 1 & 0 & \frac{\sqrt{3}}{2} \\
 0 & 0 & \frac{\sqrt{3}}{2} & 0 \\
\end{array}
\right), \, \,
%%%%%%%%%%%%%%%
J_y = \left(
\begin{array}{cccc}
 0 & \frac{-\sqrt{3} \, i } {2}   & 0 & 0 \\
 \frac{ \sqrt{3}\, i }{2} & 0 & -  i  & 0 \\
 0 &  i  & 0 & \frac{ - \sqrt{3}\, i }{2}  \\
 0 & 0 & \frac{ \sqrt{3} \, i } {2} & 0 \\
\end{array}
\right)\,,\,\,
%%%%%%%%
J_z =\frac{1}{2} \left(
\begin{array}{cccc}
 3 & 0 & 0 & 0 \\
 0 & 1 & 0 & 0 \\
 0 & 0 & -1 & 0 \\
 0 & 0 & 0 & -3 \\
\end{array}
\right)\,.
\end{align}
Here, $v_g$ denotes the magnitude of the group velocity of the quasiparticles, and we will set $v_g=1 $ for the sake of simplicity.

%%%%%%%%%%%%%%%%%%%fig 1 %%%%%%%%%%%%%%%%%%%%%%%%%%%%
\begin{figure}[t]
\subfigure[]{\includegraphics[width = 0.49 \textwidth]{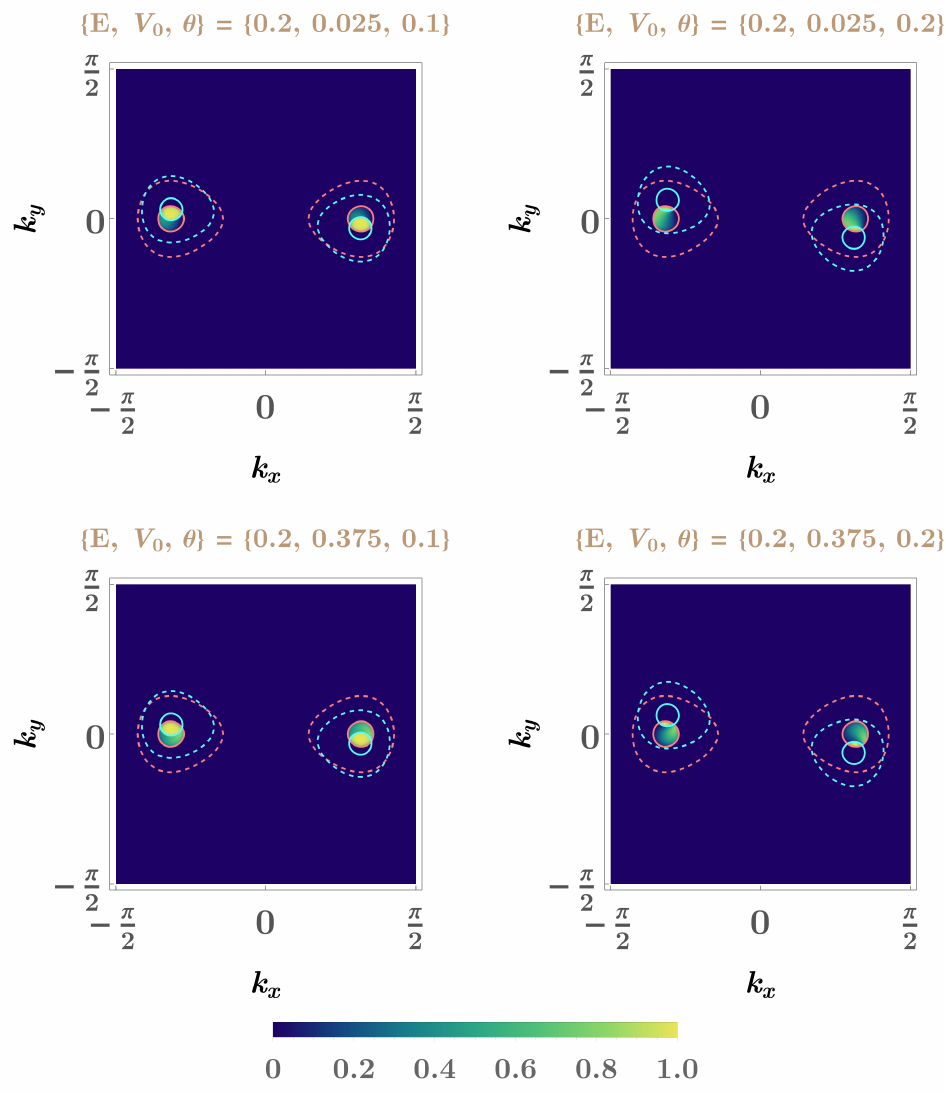}}\hspace{0.1 cm}
\subfigure[]{\includegraphics[width = 0.475 \textwidth]{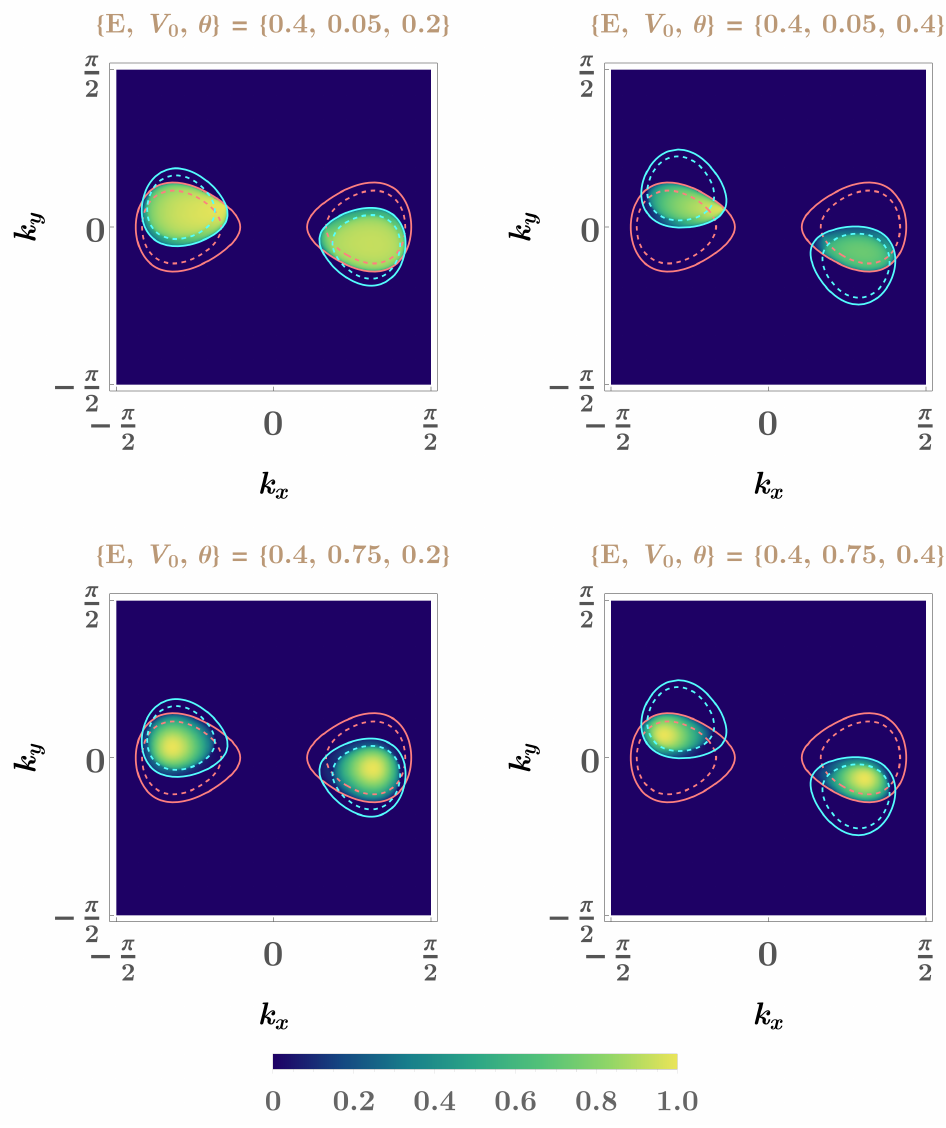}}
\caption{\label{figrsw1}
\textbf{(a) RSW semimetal} with $k_{\text{sep}}=1$: The Fermi surfaces, projected in the $k_x k_y$-plane, on the (1) left of the junction are shown by solid pink [for the band with dispersion $\mathcal{E}_{3/2}^+$) and dashed pink (for the band with dispersion $\mathcal{E}_{1/2}^+$] curves; (2) right of the junction are shown by solid cyan (for the band with dispersion $\mathcal{E}_{3/2}^{\text{sgn}(E-V_0)}$)
and dashed cyan [for the band with dispersion $\mathcal{E}_{1/2}^{\text{sgn}(E-V_0)}$] curves. Assuming the incident quasiparticles to be occupying the $\mathcal{E}_{3/2}^+$ eigenstate with energy $E$, the intensity of the transmission coefficient $\mathcal T$ is also demonstrated.
%%%%%%%%%%
\textbf{(b) bDirac semimetal} with $k_{\text{sep}}=1$ and $\zeta=0.1$:
The Fermi surfaces, projected in the $k_x k_y$-plane, on the (1) left of the junction are shown by solid pink [for the band with dispersion $\mathcal{E}_{-\zeta}^+$] and dashed pink [for the band with dispersion $\mathcal{E}_{+\zeta}^+$] curves; (2) right of the junction are shown by solid cyan [for the band with dispersion $\mathcal{E}_{-\zeta}^{\text{sgn}(E-V_0)}$] and dashed cyan [for the band with dispersion
$\mathcal{E}_{+\zeta}^{\text{sgn}(E-V_0)}$] curves. Assuming the incident quasiparticles to be occupying the
$\mathcal{E}_{-\zeta}^+$ eigenstate with energy $E$, the intensity of the transmission coefficient $\mathcal T$ is also demonstrated.
}
\end{figure}
%%%%%%%%%%%%%%%%%%%%%%%%

The energy eigenvalues take the forms:
\begin{align}
\label{eqdisrsw}
\mathcal{E}_{3/2}^s(\mathbf{k}) = s \,\frac{ 3\,|\mathbf k|} {2} \,,
\text{ and } \mathcal{E}_{1/2}^{s}(\mathbf{k}) = s \,\frac{  |\mathbf k|} {2} \,,
\text{ where } s=\pm \,,
\end{align}
demonstrating four linearly dispersing bands crossing at a point.
Here the ``$+$" and ``$-$" signs, as usual, refer to the conduction and valence bands, respectively.
The corresponding normalized eigenvectors are given by
\begin{align}
& \Psi^s_{3/2} (k_x,k_y,k_z) = \frac{1} {\mathcal{N}^s_{3/2} }
\left (
\frac{ s\,k \left( k_x^2+k_y^2+4\, k_z^2\right)+
k_z \left(3\, k_x^2+3 k_y^2+4 \,k_z^2\right)} { \left (k_x+   i  \, k_y \right )^3}
\quad
\frac{\sqrt{3} \left [ 2\, k_z \left (s\, k+k_z \right )+k_x^2+k_y^2\right ]}
{\left (k_x+  i  \,k_y\right )^2}
\quad
\frac{\sqrt{3} \left (s\, k+k_z \right )}
{k_x+  i \,k_y}
\quad 1\right)^T
\nn & \hspace{0 cm} \left(\text{for energy } \mathcal{E}_{3/2}^{s} \right)
\nn
%%%%%%%%%%%%%%%%%%%
& \text{and }
\Psi^s_{1/2}  (k_x,k_y,k_z) = \frac{1} { \mathcal{N}^s_{1/2} }
\left( \frac{ - \left( s\, k+k_z  \right )  \left (k_x-  i \, k_y \right )}
{(k_x+  i \,k_y)^2}
\quad
 \frac{2 \,k_z \left ( s\, k+k_z \right ) -k_x^2 - k_y^2}
{\sqrt{3} \left (k_x+  i \, k_y \right )^2}
\quad \frac{ s\, k+3 \, k_z}{\sqrt{3} 
\left (k_x+  i \, k_y \right )}
\quad  1\right) ^T
\left(\text{for energy } \mathcal{E}_{1/2}^{s} \right),
\end{align}
where $ k=\sqrt{k_x^2+k_y^2+k_z^2}$, and $ {1} / {\mathcal{N}^s_{3/2} }$ and $ {1} / {\mathcal{N}^s_{1/2}}$ denote the corresponding normalization factors.

The Nielsen–Ninomiya theorem \cite{nielsen-ninomiya} tells us that nodal-points harbouring chiral quasiparticles always appear in pairs. Hence, instead of considering the case of a single RSW cone,
we model the system such that two RSW cones of opposite chiralities exist. This is achieved by replacing $k_x$ with
\begin{align}
\label{eqmdef}
m= \frac{k_x^2 - k_{\text{sep}}^2}{2\,k_{\text{sep}}}
\end{align}
in Eq.~\eqref{hamrsw0} (cf. Refs.~\cite{armitage-weyl,Bovenzi_2018}). 
The modified Hamiltonian now takes the form:
\begin{align}
\label{hamrsw}
\mathcal{H}_{ \text{RSW}}(\mathbf  k) =   
m \,J_x  +  k_y\,J_y + k_z\, J_z\, ,
\end{align}
which represents two nodes with linear dispersions at the momenta
$(\eta\,k_{\text{sep}},0,0)$, with $ \eta=\pm1$ denoting the chiralities of the respective nodes.
The eigenvalues and wavefunctions are modified accordingly. 

Across the junction, on the righthand side, the two nodal points get rotated to the new positions
\begin{align}
\label{eqkw}
\left \lbrace k^\eta_{\text{sep},x}, k^\eta_{\text{sep},y}
\right  \rbrace
= \eta \,k_{\text{sep}}
\left \lbrace \cos \theta, -\sin \theta \right  \rbrace\,.
\end{align}
Let $\tilde \eta$ be the chirality of the node into which the quasiparticle is transmitted, such that
\begin{align}
\label{eqeta}
\tilde \eta = \begin{cases}
\eta &\text{ for } 0\leq \theta \leq \pi/2  \\
- \eta &\text{ for } \pi/2 <\theta \leq \pi
\end{cases}\,.
\end{align}
In this paper, we will only consider the $0\leq \theta \leq \pi/2$ cases.

In the presence of a potential step, accompanied by a rotation in the $xy$-plane, we need to consider the total Hamiltonian
\begin{align}
\mathcal{H}_{ \text{RSW}}^{tot} 
=\begin{cases} 
\mathcal{H}_{ \text{RSW}}(m, k_y,  -i\,\partial_z)
 &\text{ for }  z\leq 0 \\
\mathcal{H}_{\text{RSW}}( b^\theta_x, b^\theta_y,  -i\,\partial_z )
 + V_0 &\text{ for } z>0
\end{cases},
\end{align} 
where
\begin{align}
\label{eqbdef}
& b^\theta_x = m_\theta  \cos \theta + k_y^\theta \sin \theta \,,\quad 
 b^\theta_y =  k_y^\theta  \cos \theta -m_\theta \sin \theta\,,\quad 
m_\theta  =\frac{ \left( k_x^\theta \right)^2-k_{\text{sep}}^2} {2 \,k_{\text{sep}}} \,,\nn
& k_x^\theta = k_x \cos \theta -k_y \sin \theta \,,\quad
k_y^\theta = k_y \cos \theta + k_x \sin \theta \,.
\end{align}
We have written the $z$-dependent parts of $\mathcal{H}_{3/2}^{tot}$ in position space representation, as the translation symmetry is broken in that direction. The transverse momentum components $k_x$ and $k_y $ are conserved across the junction, as the potential is introduced only along the $z$-axis.

%%%%%%%%%%%%%%%%%%%fig 2 %%%%%%%%%%%%%%%%%%%%%%%%%%%%
\begin{figure}[t]
\subfigure[]{\includegraphics[width = 0.35 \textwidth]{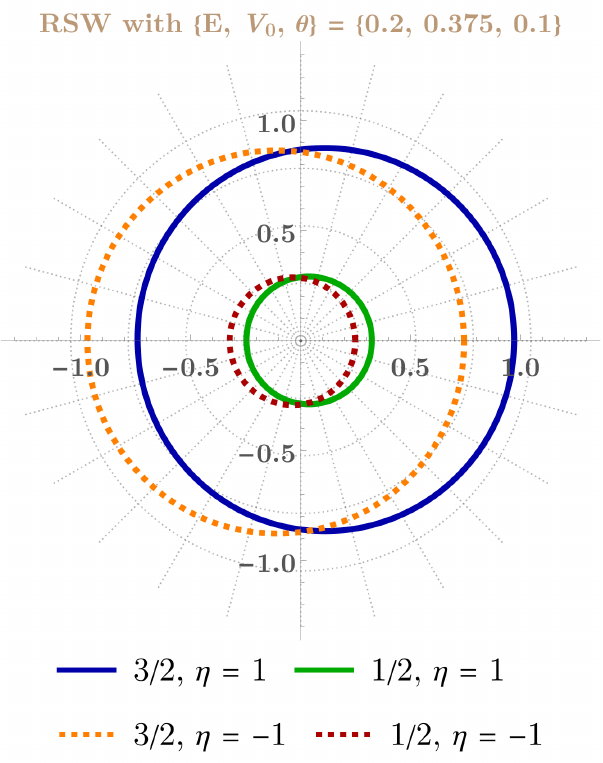}}\hspace{2 cm}
\subfigure[]{\includegraphics[width = 0.36 \textwidth]{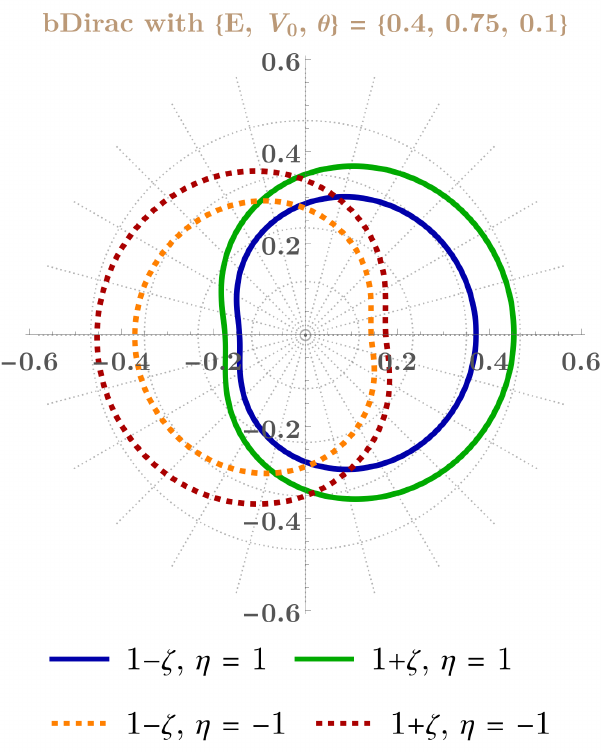}}
\caption{\label{figrsw2}
\textbf{(a) RSW semimetal} with $k_{\text{sep}}=1$: Polar plots of $\sin \theta^{out}_{3/2}$ (solid blue and dotted orange curves) and $\sin \theta^{out}_{1/2}$ (solid green and dotted red curves), with the angles defined in Eq.~\eqref{eqang1}, as functions of the azimuthal angle
$\phi^{in}$. The angle of incidence $\theta^{in}_{3/2}$ is fixed at $0.1$ (implying $\tilde \eta =\eta $).
%%%%%%%%%%
\textbf{(b) bDirac semimetal} with $k_{\text{sep}}=1$ and $\zeta=0.1$:
Polar plots of $\sin \theta^{out}_{-\zeta}$ (solid blue and dotted orange curves) and $\sin \theta^{out}_{+\zeta}$ (solid green and dotted red curves), with the angles defined in Eq.~\eqref{eqang2}, as functions of the azimuthal angle $\phi^{in}$. The angle of incidence $\theta^{in}_{-\zeta}$ is fixed at $0.1$ (implying $\tilde \eta =\eta $).
}
\end{figure}

We will compute the transmission coefficient $\mathcal T(E,V_0,\theta)$ when the incident wavefunction is in the $\Psi^+_{3/2} $ eigenstate and $E = 3\,\sqrt{m^2+k_y^2 +k_z^2}  \, /2$. For this purpose,
a scattering state can be constructed as
\begin{align}
 & \hspace{6 cm} \Psi_{k_x,k_y,E} (z)=  \begin{cases}   
 \phi_{L} & \text{ for } z \leq 0  \\
  \phi_{R} &  \text{ for } z > 0 
\end{cases} \,,\nonumber \\
%%%%%%%%%%%%%%%%%%%%%%%%%%%%%%%%%%%%%%%%
&  \phi_{L} =   
 \Psi^+_{3/2}  (m,k_y,k_{z_1})\,e^{  i \, k_{z_1}  z}
 + r_1 \, \Psi^+_{3/2} (m ,k_y,- k_{z_1})\,e^{-  i \, k_{z_1} z}
 + r_2 \, \Psi^+_{1/2} (m ,k_y,- k_{z_2})\,e^{-  i \, k_{z_2} z}\, ,\nonumber \\
%%%%%%%%%%%%%%%%%%%%%%%%%%%%%%%%%%%%%%%%%%%%5
&  \phi_{R}  =  
t_1 \, \Psi^{\text{sgn}(E-V_0)}_{3/2} ( b^\theta_x, b^\theta_y, {\tilde k}_{z_1})
\,e^{  i \, {\tilde k}_{z_1} z}
 + t_2 \, \Psi^{\text{sgn}(E-V_0)}_{1/2} (b^\theta_x , b^\theta_y , {\tilde k}_{z_2})
 \,e^{  i \, {\tilde k}_{z_2} z}\, ,\nonumber \\
%%%%%%%%%%%%%%%%%%%%%%%%%%%%%%%%% 
& k_{z_1} \equiv k_z  = \sqrt{\left(\frac{E}{3/2 }\right)^2-m^2-k_y^2}
 \,, \quad
%%%%%%%%%%%%%%%%%%%%%%
k_{z_2} = \sqrt{\left(\frac{E} {1/2}\right)^2-m^2-k_y^2}\,,\nn
%%%%%%%%%%%%%%%%%%%%%%%%%%%%%%%%% 
& {\tilde k}_{z_1}   = \text{sgn}(E-V_0)\,
\sqrt{\left(\frac{E-V_0}{ 3/2 }\right)^2
-\left(b^\theta_x \right) ^2- \left( b^\theta_y\right)^2}
 \,, \quad
%%%%%%%%%%%%%%%%%%%%%%
{\tilde k}_{z_2}\ = \text{sgn}(E-V_0) \, \sqrt{\left(\frac{E-V_0} {1/2 }\right)^2
-\left(b^\theta_x \right) ^2- \left( b^\theta_y\right)^2}\,.
\end{align}
Here, $r_1$ and $ r_2$ represent the reflection amplitudes in the eigenstates $\mathcal{E}^+_{3/2}$
and $\mathcal{E}^+_{1/2}$, respectively. On the other hand, $t_1$ and $ t_2$ represent the transmission amplitudes in the channels $\mathcal{E}^{\text{sgn}(E-V_0)}_{3/2}$
and $\mathcal{E}^{\text{sgn}(E-V_0)}_{1/2}$, respectively. We note that $t_1$ (or $t_2$) is defined only when ${\tilde k}_{z_1}$ (or ${\tilde k}_{z_2}$) is real, as a decaying mode does not contribute to transmission coefficient.

Imposing the continuity of the wavefunction at the junction $z=0$, we get four equations from the four components of the wavefunction. These four linearly independent equations are solved to determine the four unknown coefficients $r_1$, $r_2$, $t_1$, and $t_2$. Because the group velocities are different on the two sides of the junction, we need to use the relation
[cf. Eq.~\eqref{eqtT}]
\begin{align}
& \mathcal T(E,V_0,\theta)
 = \sum_{\gamma= 1, \,2} \mathcal T_\gamma (E,V_0,\theta) \,, \nn & \nn
%%%%%%%%%%%%%%%%
 \mathcal T_1(E,V_0,\theta)
& =  \left | \sqrt{
\frac{
{\tilde k}_{z_1}
/
\sqrt{ \left(b^\theta_x \right) ^2+ \left(b^\theta_y \right) ^2+  {\tilde k}_{z_1}^2}
}
{ {k}_{z_1} /
\sqrt{m^2 + k_y^2 +  {k}_{z_1}^2 }
}}
\,\,\,t_1  \right |^2, \quad
%%%%%%%%%%%%%%%%%%%%%%%%%%%%
 \mathcal T_2(E,V_0,\theta) =
\left | \sqrt{
\frac{
 {\tilde k}_{z_2}
/
\sqrt{ \left(b^\theta_x \right) ^2+ \left(b^\theta_y \right) ^2+  {\tilde k}_{z_2}^2}
}
{ 3  \, {k}_{z_1} /
\sqrt{m^2 + k_y^2 +  {k}_{z_1}^2 }
}}
\,\,\,t_2  \right |^2  .
\end{align}
%%%%%%%%%%%%%
Fig.~\ref{figrsw1}(a) shows the transmission spectrum of the system for various values of the parameters.
We find that $\mathcal T$ is nonzero in the area where the projections of the Fermi surfaces from the two sides of the junction overlap. We note that near each nodal-point, we have two Fermi surfaces around each node due to the four-band structure. Since the incident quasiparticle is assumed to be confined to the inner Fermi surface of each node, the transmission spectrum is bounded by the inner Fermi surfaces of the lefthand and righthand sides on the junction. We have also checked that $t_2$ can be nonzero in generic cases.

Since the transmission of a quasiparticle takes
place in the vicinity of a node, its dispersion relation is approximately linear in the deviation of the momentum from the position of the node in consideration. Remembering this, the group velocity of the incident quasiparticles, propagating in the eigenstate $\mathcal{E}_{3/2}^+ $ with energy $E$,
is given by $ \boldsymbol{v}^{in}_{3/2} = \partial_{\mathbf k} \mathcal{E}_{3/2}^+$. In component form, 
we get 
\begin{align}
\boldsymbol{v}^{in}_{3/2} \equiv \left \lbrace v_x^{in,3/2}, v_y^{in}, v_z^{in}  \right  \rbrace =
\frac{3/2} {E} 
\left \lbrace k_x-\eta\,k_{\text{sep}}, k_y, k_z \right \rbrace .
\end{align}
Using Eqs.~\eqref{eqkw} and \eqref{eqeta}, we determine that the group velocity of the outgoing quasiparticles
to be
\begin{align}
& \boldsymbol{v}^{out}_{3/2} \equiv
\left \lbrace v_x^{out,3/2}, v_y^{out}, v_z^{out}  \right  \rbrace =
\frac{3/2} {E-V_0} \left \lbrace 
k_x-{\tilde \eta}\,k_{\text{sep},x}^{\tilde \eta},
 k_y -{\tilde \eta}\,k_{\text{sep},y}^{\tilde \eta}, 
{\tilde k}_{z_1} \right \rbrace
%%%%%%%%%%%%%%%%%%%%%%
\\ & \text{or }
\boldsymbol{v}^{out}_{1/2} \equiv
\left \lbrace v_x^{out,3/2}, v_y^{out}, v_z^{out}  \right  \rbrace =
\frac{1/2} {E-V_0} \left \lbrace 
k_x-{\tilde \eta}\,k_{\text{sep},x}^{\tilde \eta},
 k_y -{\tilde \eta}\,k_{\text{sep},y}^{\tilde \eta}, {\tilde k}_{z_2} 
 \right \rbrace ,
\end{align}
depending on the mode occupied. We now parameterize these group velocities in terms of angular variables as follows (remembering that we have set $v_g=1$):
\begin{align}
& \boldsymbol{v}^{in}_{3/2} = 
\left \lbrace 
\sin \theta^{in}_{3/2} \sin \phi^{in}, 
\sin \theta^{in}_{3/2} \cos \phi^{in}, 
\cos \theta^{in}_{3/2} \right \rbrace, \nn 
%%%%%%%%%%%
& \boldsymbol{v}^{out}_{3/2} = 
\left \lbrace 
\sin \theta^{out}_{3/2} \sin \phi^{out}, 
\sin \theta^{out}_{3/2} \cos \phi^{out}, 
\cos \theta^{out}_{3/2} \right \rbrace, \quad
%%%%%%
\boldsymbol{v}^{out}_{1/2} = 
\left \lbrace 
\sin \theta^{out}_{1/2} \sin \phi^{out}, 
\sin \theta^{out}_{1/2} \cos \phi^{out}, 
\cos \theta^{out}_{1/2} \right \rbrace
\end{align}
such that the set $\left \lbrace \theta^{in}_{3/2}, \theta^{out}_{3/2}, \theta^{out}_{1/2} \right \rbrace$ denotes the angles that the velocity vectors make with the normal to the interface, and the corresponding azimuthal angles in the $xy$-plane are denoted by $\left \lbrace \phi^{in}, \phi^{out} \right \rbrace $. Each of the incident angles
can lie between $0$ and $\pi/2$, whereas each of the azimuthal angles can lie between $0$ and $\pi$.

According to our conventions, we have
\begin{align}
&  k_z^{(3/2)} \equiv k_z =\frac{E} {3/2} \cos \theta^{in}_{3/2}\,, \quad 
k_y =\frac{E} {3/2} \sin \theta^{in}_{3/2} \cos \phi^{in} \,,\quad
k_x = \frac{E} {3/2} \sin \theta^{in}_{3/2} \sin \phi^{in} +\eta \, k_{\text{sep}}\,,
%%%%%%%%%%%%%%%
\nn & k_z^{(1/2)} = \sqrt{
\left( \frac{E} {1/2} \right)^2 
- \left(k_x -\eta\, k_{\text{sep}} \right)^2 -k_y^2
} \,,\nn
%%%%%%%
& k_x^\theta = k_x \cos \theta -k_y \sin \theta \,,\quad
k_y^\theta = k_y \cos \theta + k_x \sin \theta\,,
%%%%%%%%%%%%%%
\quad  b^\theta_x =k_x - {\tilde\eta } \,k_{\text{sep}} \cos \theta \,,\quad 
 b^\theta_y =k_y + {\tilde\eta } \,k_{\text{sep}} \sin \theta \,,\nn
%%%%%%%%%%
&  {\tilde k}_z^{(3/2)} =\text{sgn}(E-V_0)  \,
\sqrt{ 
\left( \frac{E-V_0} {3/2}  \right)^2-\left( b^\theta_x \right)^2
-\left( b^\theta_y \right)^2
} \,,\quad 
%%%%%%%%%
{\tilde k}_z^{(1/2)} =\text{sgn}(E-V_0) \,
\sqrt{ \left( \frac{E-V_0} {1/2} \right)^2 -\left( b^\theta_x \right)^2
-\left( b^\theta_y \right)^2 }\,.
\end{align}
%%%%%%%%%%%%%%
Using the above, the angles of refraction, for transmission into the $\mathcal{E}_{3/2}^{\text{sgn} (E-V_0)}$ and
$\mathcal{E}_{1/2}^{\text{sgn} (E-V_0)}$ bands, are given by
\begin{align}
\label{eqang1}
\theta^{out}_{3/2} =
\cos ^{-1}\left(\frac{ 3\,{\tilde k}_z^{(3/2)}}
{  2 \left| E-V_0\right| }\right) \text{ and }
\theta^{out}_{1/2} =
\cos ^{-1}\left(\frac{  {\tilde k}_z^{(1/2)}}
{  2 \left| E-V_0\right| }\right),
\end{align}
respectively. We note that the angle reflection into the $\mathcal{E}_{1/2}^{+}$ mode is given by
$\theta^{in}_{1/2} = \cos ^{-1}\left(\frac{  {k}_z^{(1/2)}}
{  2 \, E }\right)$.
The behaviour of $\sin \theta^{out}_{3/2}$ and $\sin \theta^{out}_{1/2}$, as functions of the incident azimuthal angle $\phi^{in}$ in the $k_x k_y$-plane, are shown in Fig.~\ref{figrsw2}(a).
Clearly, we find that the angles of refraction are different from the angle of incidence, resulting from the presence of the tilt $\theta$.

%%%%%%%%%%%%%%%%%%% fig 4 %%%%%%%%%%%%%%%%%%%%%%%%%%%%
\begin{figure}[t]
\subfigure[]{\includegraphics[width = 0.45 \textwidth]{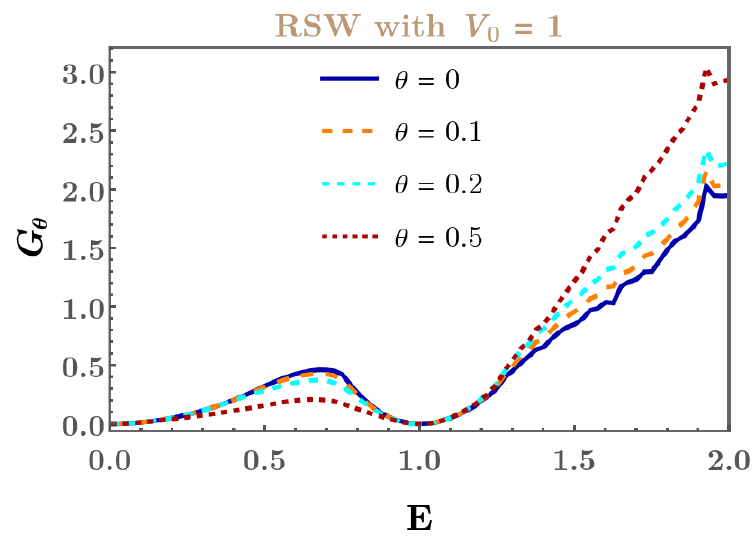}}\hspace{1 cm}
\subfigure[]{\includegraphics[width = 0.44 \textwidth]{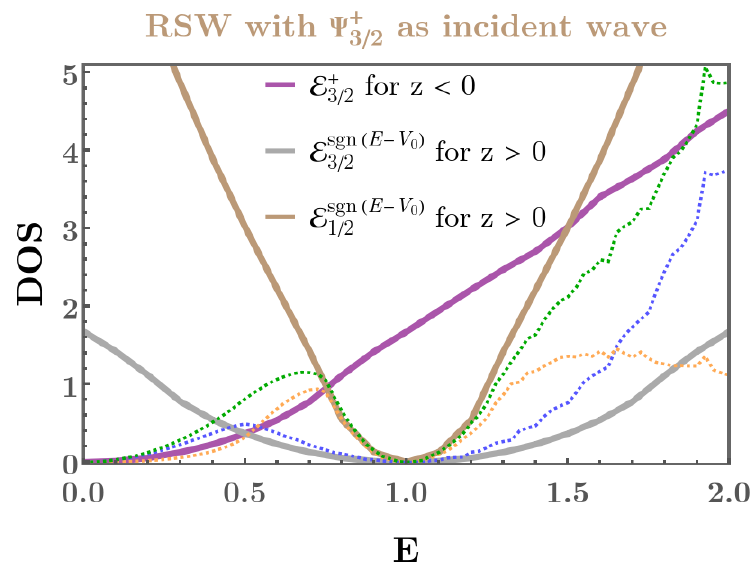}}
\caption{\label{figrsw3}
\textbf{RSW semimetal} with $ k_{\text{sep}}=1 $ and $V_ 0 =  1 $:
\textbf{(a)} The curves show the behaviour of the zero-temperature conductance per unit surface area [in units of
$ e^2/(2\,\pi)$] versus $ E $ for different values of the rotation angle $\theta$ (indicated in the plotlegend).
%%%%%%%%%%
\textbf{(b)} The curves show the DOS [in units of $(2\,\pi)^{-2}$] for the regions on the (I) left of the junction (solid light-purple curve); and (II) right of the junction (solid light-gray and light-brown curves correspond to the eigenstates $ \Psi^{\text{sgn}(E-V_0)}_{3/2} $ and $ \Psi^{\text{sgn}(E-V_0)}_{1/2} $ , respectively). The dotted blue, orange, and green curves represent $G_\theta^{(1)}$, $G_\theta^{(2)}$, and $ G_\theta = G_\theta^{(1)} + G_\theta^{(2)}$, respectively, for $\theta =0 $ and in units of
$ 0.4 \, e^2/(2\,\pi)$.
}
\end{figure}
%%%%%%%%%%%%%%%%%%%%%%

We also evaluate the zero-temperature conductance $G_\theta$ per unit surface area, using the coordinates $k_z =\frac{2 \,E}{3}  \cos {\tilde \theta}$,  $k_y=\frac{2\, E}{3}  \sin  {\tilde \theta} \cos  \phi $, $m=\frac{2\,E}{3} \sin  {\tilde \theta} \sin  \phi $,
and $ k_x= \pm \sqrt{k_{\text{sep}}\, (k_{\text{sep}}+2 \,m)}$.
The expression for $G$ is explicitly given by \cite{blanter-buttiker}
\begin{align}
G_\theta(E,V_0)  
&= \frac{e^2}{2\,\pi} \int dk_x \,dk_y \,\mathcal T(E,V_0,\theta)
= \sum_{\gamma = 1, \, 2 } G^{(\gamma)}_\theta(E,V_0)  \,,\nn  
%%%%%%%%%%%%
 G^{(\gamma)}_\theta(E,V_0) & 
= \frac{e^2}{2\,\pi}
\int dk_x \,dk_y\,\mathcal T_\gamma (E,V_0,\theta)
%%%%%%% 
 = \frac{e^2}{2\,\pi}
 \int_{ \tilde \theta=0 }^{\pi/2} \int_{\phi=0}^{2\,\pi}
 d{\tilde \theta} \,d\phi\,
\left| 
\frac{2 \,E^2 \,
\sqrt{k_{\text{sep}}} \, \sin (2 \tilde{\theta })
}
{3 \,\sqrt{3} \,\sqrt{4 \,E \sin \phi 
\sin \tilde{\theta }  + 3\, k_{\text{sep}}}}
 \right|\,
\mathcal T_\gamma (E,V_0,\theta) \,.
%\,, \text{ where } \gamma \in \lbrace1, \,2 \rbrace \,.
\end{align}
%%%%%%
Fig.~\ref{figrsw3}(a) shows the $ G_\theta $ curves as functions of $E$, with $V_0$ fixed to the value of unity, for some representative values of the remaining parameters. Fig.~\ref{figrsw3}(b) shows the density of states (DOS) for the regions on the (I) left of the junction (solid light-purple curve); and (II) right of the junction (solid light-gray and light-brown curves correspond to the eigenstates $ \Psi^{\text{sgn}(E-V_0)}_{3/2} $ and $ \Psi^{\text{sgn}(E-V_0)}_{1/2} $ , respectively).
The DOS curves provide intuition for interpreting the behaviour of the conductivity curves as explained below:
%%%%%%%%%%%%%%%%%%
\begin{itemize}

\item
For $E = 0 $, the Fermi energy on the left of the junction cuts the band-crossing point where the DOS is zero, resulting in $ G_\theta = 0 $ as well. For the regime $E<V_0$, the DOS in $z<0$ increases and the DOS in $z > 0$ decreases. As we increase $E$ from zero, each transmission probability $\mathcal T_{1} $ ($\mathcal T_{2} $) increases monotonically until the DOS for $z<0$ equals the DOS of the $\mathcal{E}^{\text{sgn}(E-V_0)}_{3/2}$ ($\mathcal{E}^{\text{sgn} (E-V_0) }_{1/2}$) band. Thus each of the two $G_\theta^{(\gamma)} $-curves reaches a local maximum --- (I) for $ G_\theta^{(1)} $ [indicated by the dotted blue curve in Fig.~\ref{figrsw3}(b) taking $\theta=0$], this happens at $ E = 0.5 $ where the solid light-purple curve intersects the light-gray curve; (II) for $G_\theta^{(2)} $ [indicated by the dotted orange curve in Fig.~\ref{figrsw3}(b) taking $\theta=0$], this happens at $ E = 0.75 $ where the solid light-purple curve intersects the light-brown curve.
%%%%
The sum $ G_\theta = G_\theta^{(1)} + G_\theta^{(2)}$ [indicated by the dotted green curve in Fig.~\ref{figrsw3}(b) taking $\theta=0$] thus reaches a local maximum somewhere between $ E = 0.5 $ and $ E = 0.75 $, which is what is observed in Fig. \ref{figrsw3}(a).
%%%%%%%%%%

\item
As $E$ is cranked up further beyond this point, $G_\theta $ monotonically decreases until it reaches the local minimum value of zero at $E = V_0 $, because the DOS for $z>0$ becomes zero there.

\item
For $E  > V_0 $, the DOS in either region increases, leading to a continuous increase of $G_\theta $. We observe another local maximum for $G_\theta^{(2)} $ at $ E =1.5 $, when the the solid light-purple curve intersects the light-brown curve a second time. This behaviour arises only arises because of four-bands, and is not seen for a two-band system (e.g., the Weyl semimetal cosidered in Ref.~\cite{tilted_jn}). Furthermore, this second crossing occurs only because the incident particle is occupying the band with the higher energy eigenvalue represented by the inner cones in the dispersion.

\item The relative behaviour of the individual $G_\theta$-curves for distinct values of $ \theta$ can be understood from the fact that although the DOS for $z>0$ does not depend on $\theta $ (as we obtain it by integrating over the $k_x k_y$-plane), the momentum-space region, where the transmission coefficient $\mathcal T$ is vanishing, broadens with increasing $\theta$, resulting in a greater mismatch between the scattering states on the corresponding Fermi surface projections on the two sides of the junction [cf. Fig.~\ref{figrsw1}(a)]. We note that the region around the minimum at $ E = V_0 $ becomes flatter / broader with increasing $ \theta $ because of the interplay of two effects --- (I) as $|E - V_0|$ becomes smaller, the Fermi surface projection in the $ z>0$ region gets smaller and smaller; and (II) the change in the location of the nodal points can be measured by the vector $ k _{\text{sep}}
\left \lbrace 1 - \cos \theta, \sin \theta \right  \rbrace$ [cf. Eq.~\eqref{eqkw}], which thus quantifies the amount of mismatch between the momenta of the scattering states in the $k_x k_y$-plane. The mismatch in momenta increases as the value of $ \theta $ increases in the range $[0,\, \pi/2]$. The combination of a smaller Fermi surface projection and a greater mismatch in the perpedicular-to-the-junction momentum therefore leads to less transmission, which reflects in the broadening of the zero conductivity area around $E = V_0 $ with increasing values of $\theta $.

\end{itemize}

%%%%%%%%%%%%%%%%%%%%%%%%%%%%%%%%%%5
\subsection{Birefringent semimetal}

The Hamiltonian describing the pseudospin-3/2 birefringent quasiparticles (abbreviated here as bDirac) is given by \cite{malcolm,malcolm-bitan}
\begin{align}
\label{hambiref1}
& \mathcal{H}_{  \text{bDirac}}^{1\text{node}}(\mathbf  k) 
 = H_1(\mathbf k) +H_2(\mathbf k)\,, 
\quad H_1(\mathbf k) = v\sum_{j=x,y,z}\Gamma_{j0}\,k_j\,, 
\quad H_2(\mathbf k)= v\sum_{j=x,y,z}\zeta\,\Gamma_{0 j}\,k_j \,,\nn 
& \Gamma_{j 0} =\sigma_j \otimes \sigma_0\,,\quad
\Gamma_{0 j} =\sigma_0 \otimes \sigma_j\,,
\end{align}
where $\sigma_0$ is the $2\times 2$ identity matrix. The four eigenvalues are given by
\begin{align}
\label{eqevbiref}
\mathcal{E}^s_{\beta \zeta} = s\, v \left( 1 +\beta\, \zeta \right) |\mathbf k|\,,
\quad \text{ where } s=\pm \text{ and }\beta=\pm.
\end{align} 
Hence, the emergent quasiparticles have birefringent spectra with linear dispersions. Here $\zeta$  is the birefringence parameter, with $|\zeta| < 1$.

In the following, we rescale the spatial momenta such that $v$ is equal to one.
The normalized eigenvectors are given by
\begin{align}
& \Psi^s_{-\zeta} (k_x,k_y,k_z) = \frac{1} {\mathcal{N}^s_{-\zeta} }
\left (
1-\frac{2 \,k_x}
{k_x+i \,k_y}
\quad
\frac{\frac{s \,k}  {1-\zeta }+k_z}
{k_x+i\, k_y},\frac{k_z
-\frac{s \,k}{1-\zeta }}{k_x+i \,k_y}
\quad 1 \right )^T 
\left(\text{for energy } \mathcal{E}^s_{-\zeta} \right)
\nn
%%%%%%%%%%%%%%%%%%%
\text{and } &\Psi^s_{+\zeta}(k_x,k_y,k_z)  = \frac{1} { \mathcal{N}^s_{+\zeta} }
\left (
\frac{k_x^2+k_y^2+ 2 \,k_z \left(\frac{s\, k}{ 1+ \zeta }+k_z\right)}
{\left(k_x+i \,k_y\right){}^2},\frac{\frac{s\, k} { 1+ \zeta }+k_z}
{k_x+i \,k_y}
\quad \frac{\frac{s\, k} { 1+ \zeta }+k_z} {k_x+i \,k_y}
\quad 1
\right )^T 
\left(\text{for energy } \mathcal{E}^s_{+\zeta} \right),
\end{align}
where $ k=\sqrt{k_x^2+k_y^2+k_z^2}$, and $ {1}/ {\mathcal{N}^s_{-\zeta} }$ and ${1} /{\mathcal{N}^s_{+\zeta}}$ represent the corresponding normalization factors.

Similar to the RSW case, we consider a more generic system with two cones of opposite chiralities by employing the Hamiltonian
\begin{align}
\label{hambiref2}
\mathcal{H}_{  \text{bDirac}}(\mathbf  k) =   
m \left( \Gamma_{x0} + \zeta\,\Gamma_{0x}\right)
+ k_y \left( \Gamma_{x0} + \zeta\,\Gamma_{0x}\right)
+k_z \left( \Gamma_{z0} + \zeta\,\Gamma_{0z}\right),
\end{align}
with $m$ given by Eq.~\eqref{eqmdef}.
Hence, this configuration represents two nodes with linear dispersions at the momenta
$(\eta\,k_{\text{sep}},0,0)$, with $ \eta=\pm1$ denoting the chiralities of the respective nodes.

In the presence of a potential step, accompanied by a rotation in the $xy$-plane, we need to consider the total Hamiltonian
\begin{align}
\mathcal{H}_{  \text{bDirac}}^{tot} 
=\begin{cases} 
\mathcal{H}_{  \text{bDirac}}(m, k_y,  -i\,\partial_z)
 &\text{ for }  z\leq 0 \\
\mathcal{H}_{ \text{bDirac}}( b^\theta_x, b^\theta_y,  -i\,\partial_z )
 + V_0 &\text{ for } z>0
\end{cases},
\end{align} 
with $\lbrace  k^\theta_x, k^\theta_y, b^\theta_x, b^\theta_y\rbrace $ defined by Eq.~\eqref{eqbdef}.

We will compute the transmission coefficient $\mathcal T(E,V_0,\theta)$ when the incident wavefunction is in the
$\Psi^+_{-\zeta} $ eigenstate, where $E = \sqrt{m^2+k_y^2 +k_z^2}  \, / (1-\zeta)$. For this purpose,
we define the scattering state as
\begin{align}
 & \hspace{6 cm} \Psi_{k_x,k_y,E} (z)=  \begin{cases}   
 \phi_{L} & \text{ for } z \leq 0  \\
  \phi_{R} &  \text{ for } z > 0 
\end{cases} \,,\nonumber \\
%%%%%%%%%%%%%%%%%%%%%%%%%%%%%%%%%%%%%%%%
&  \phi_{L} =   
 \Psi^+_{-\zeta}  (m,k_y,k_{z_1})\,e^{  i \, k_{z_1}  z}
 + r_1 \, \Psi^+_{-\zeta} (m ,k_y,- k_{z_1})\,e^{-  i \, k_{z_1} z}
 + r_2 \, \Psi^+_{+ \zeta} (m ,k_y,- k_{z_2})\,e^{-  i \, k_{z_2} z}\, ,\nonumber \\
%%%%%%%%%%%%%%%%%%%%%%%%%%%%%%%%%%%%%%%%%%%%5
&  \phi_{R}  =  
t_1 \, \Psi^{\text{sgn}(E-V_0)}_{-\zeta} ( b^\theta_x, b^\theta_y, {\tilde k}_{z_1})
\,e^{  i \, {\tilde k}_{z_1} z}
 + t_2 \, \Psi^{\text{sgn}(E-V_0)}_{+ \zeta} (b^\theta_x , b^\theta_y , {\tilde k}_{z_2})
 \,e^{  i \, {\tilde k}_{z_1} z}\, ,\nonumber \\
%%%%%%%%%%%%%%%%%%%%%%%%%%%%%%%%% 
& k_{z_1} \equiv k_z  = \sqrt{\left(\frac{E}{1-\zeta }\right)^2-m^2-k_y^2}
 \,, \quad
%%%%%%%%%%%%%%%%%%%%%%
k_{z_2} = \sqrt{\left(\frac{E}{1+\zeta }\right)^2-m^2-k_y^2}\,,\nn
%%%%%%%%%%%%%%%%%%%%%%%%%%%%%%%%% 
& {\tilde k}_{z_1}   = \text{sgn}(E-V_0)\,
\sqrt{\left(\frac{E-V_0}{1-\zeta }\right)^2
-\left(b^\theta_x \right) ^2- \left( b^\theta_y\right)^2}
 \,, \quad
%%%%%%%%%%%%%%%%%%%%%%
{\tilde k}_{z_2}\ = \text{sgn}(E-V_0) \, \sqrt{\left(\frac{E-V_0}{1+\zeta }\right)^2
-\left(b^\theta_x \right) ^2- \left( b^\theta_y\right)^2}\,.
\end{align}
Here, $r_1$ and $ r_2$ represent the reflection amplitudes in the eigenstates $\mathcal{E}^+_{-\zeta}$
and $\mathcal{E}^+_{+\zeta}$, respectively. Similarly, $t_1$ and $ t_2$ represent the transmission amplitudes in the channels $\mathcal{E}^{\text{sgn}(E-V_0)}_{-\zeta}$
and $\mathcal{E}^{\text{sgn}(E-V_0)}_{+\zeta}$, respectively. We must also remember that $t_1$ (or $t_2$) is defined only when ${\tilde k}_{z_1}$ (or ${\tilde k}_{z_2}$) is real, as a decaying mode does not contribute to transmission coefficient.

Imposing the continuity of the wavefunction at the junction $z=0$, we get four equations from the four components of the wavefunction. These four linearly independent euqations are solved to determine the four unknown coefficients $r_1$, $r_2$, $t_1$, and $t_2$. Accounting for the difference in group velocities on the two sides of the junction, we obtain [cf. Eq.~\eqref{eqtT}]
\begin{align}
& \mathcal T(E,V_0,\theta)
  = \sum_{\gamma= 1, \,2} \mathcal T_\gamma (E,V_0,\theta) \,, \nn & \nn
%%%%%%%%%%%%%%%%%%%%
 \mathcal T_1 (E,V_0,\theta) & = \left | \sqrt{
\frac{
{\tilde k}_{z_1}
/
\sqrt{ \left(b^\theta_x \right) ^2+ \left(b^\theta_y \right) ^2+  {\tilde k}_{z_1}^2}
}
{ {k}_{z_1} /
\sqrt{m^2 + k_y^2 +  {k}_{z_1}^2 }
}}
\,\,\,t_1  \right |^2, \quad
%%%%%%%%%%%%%%%%%%%%%%%55
\mathcal T_2 (E,V_0,\theta) = \left | \sqrt{
\frac{
(1+\zeta )  \, {\tilde k}_{z_2}
/
\sqrt{ \left(b^\theta_x \right) ^2+ \left(b^\theta_y \right) ^2+  {\tilde k}_{z_2}^2}
}
{(1-\zeta )  \, {k}_{z_1} /
\sqrt{m^2 + k_y^2 +  {k}_{z_1}^2 }
}}
\,\,\,t_2  \right |^2 .
\end{align}
%%%%%%%%%%%%%
We show some representative spectra in Fig.~\ref{figrsw1}(b).
Similar to the RSW case, $\mathcal T$ is nonzero in the area where the projections of the Fermi surfaces from the two sides of the junction overlap. Here also we have two Fermi surfaces around each node due to the four-band structure. But the incident quasiparticle in this case is assumed to be confined to the outer Fermi surface of each node. Hence, the transmission spectrum is bounded by the outer Fermi surfaces of the lefthand and righthand sides on the junction. We have also checked that $t_2$ can be nonzero in generic cases.

We determine the angles of refraction following the same method as described for the RSW case. For this case, we use the conventions
%%%%%%%%%%%%%%%%%%%%%%%%%%%%%%%
\begin{align}
&  k_z^{(-\zeta)} \equiv k_z =\frac{E} {1-\zeta} \cos \theta^{in}_{-\zeta}\,, \quad 
k_y =\frac{E} {1-\zeta} \sin \theta^{in}_{-\zeta} \,\cos \phi^{in} \,,\quad
k_x = \frac{E} { 1-\zeta} \sin \theta^{in}_{-\zeta} \,\sin \phi^{in} +\eta \, k_{\text{sep}}\,,
%%%%%%%%%%%%%%%
\nn & k_z^{(+ \zeta)} = \sqrt{
\left( \frac{E} { 1+ \zeta} \right)^2 
- \left(k_x -\eta\, k_{\text{sep}} \right)^2 -k_y^2
} \,,
%%%%%%%
\quad  k_x^\theta = k_x \cos \theta -k_y \sin \theta \,,\quad
k_y^\theta = k_y \cos \theta + k_x \sin \theta\,,
%%%%%%%%%%%%%%
\nn  &
b^\theta_x =k_x - {\tilde\eta } \,k_{\text{sep}} \cos \theta \,,\quad 
 b^\theta_y =k_y + {\tilde\eta } \,k_{\text{sep}} \sin \theta \,,\quad
%%%%%%%%%%
 {\tilde k}_z^{(\beta \zeta)} =\text{sgn}(E-V_0)  \,
\sqrt{ 
\left( \frac{E-V_0} {1 + \beta \,\zeta}  \right)^2-\left( b^\theta_x \right)^2
-\left( b^\theta_y \right)^2 } \,.
\end{align}
%%%%%%%%%%%%%%
Here, $\left \lbrace \theta^{in}_{\beta \zeta}, \theta^{out}_{\beta \zeta} \right \rbrace$ denote the angles that the velocity vectors make with the normal to the interface, and the corresponding azimuthal angles in the $xy$-plane are denoted by $\left \lbrace \phi^{in}, \phi^{out} \right \rbrace $.

Using the above, the angle of refraction for transmission into the $\mathcal{E}_{\beta \zeta}^{\text{sgn} (E-V_0)}$
band is given by
\begin{align}
\label{eqang2}
\theta^{out}_{\beta \zeta} =
\cos ^{-1}\left(\frac{(1 +\beta \, \zeta )\,{\tilde k}_z^{\beta \zeta}}
{\left| E-V_0\right| }\right).
\end{align}
We note that the angle reflection into the $\mathcal{E}_{+\zeta}^{+}$ mode is given by
$\theta^{in}_{+\zeta} = \cos ^{-1}\left(
\left( 1+\zeta \right) \, {k}_z^{(+\zeta)} / E \right)$. The behaviour of $\sin \theta^{out}_{\beta \zeta}$, as a function of the incident azimuthal angle $\phi^{in}$ in the $k_x k_y$-plane, is shown in Fig.~\ref{figrsw2}(b).

%%%%%%%%%%%%%%%%%%%fig 5 %%%%%%%%%%%%%%%%%%%%%%%%%%%%
\begin{figure}[t]
\subfigure[]{\includegraphics[width = 0.46 \textwidth]{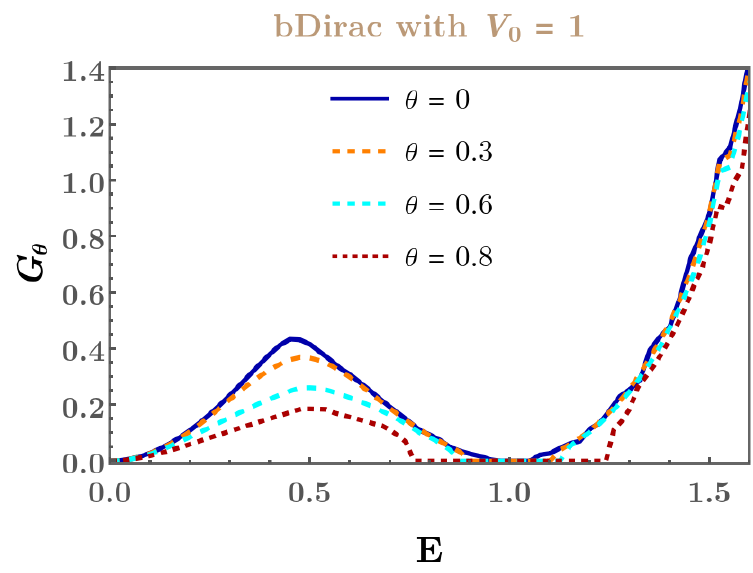}} \hspace{1 cm}
\subfigure[]{\includegraphics[width = 0.44 \textwidth]{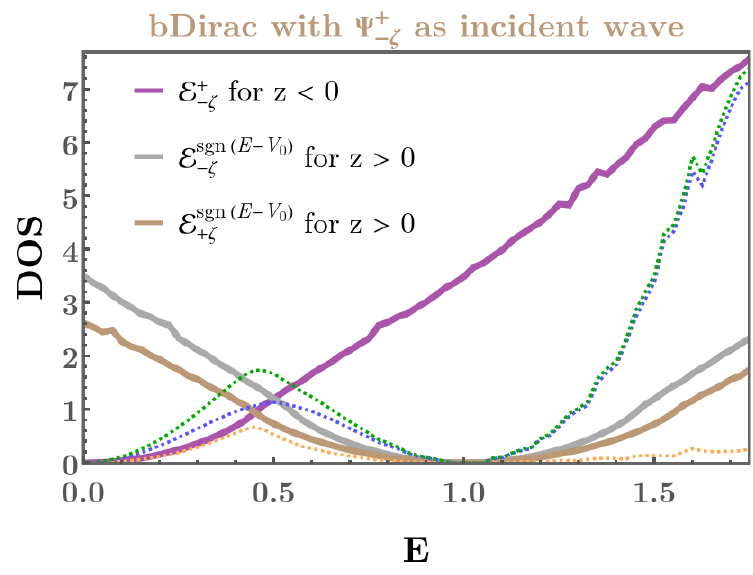}}
\caption{\label{figbdir3}
\textbf{bDirac semimetal} with $ k_{\text{sep}}=1 $, $\zeta=0.1$, and $V_ 0 = 1 $:
\textbf{(a)} The curves show the behaviour of the zero-temperature conductance per unit surface area [in units of $ e^2/(2\,\pi)$] versus $ E $ for different values of the rotation angle $\theta$ (indicated in the plotlegend).
%%%%%%%%%%
\textbf{(b)} The curves show the DOS  [in units of $(2\,\pi)^{-2}$] for the regions on the (I) left of the junction (solid light-purple curve); and (II) right of the junction (solid light-brown and light-gray curves correspond to the eigenstates $ \Psi^{\text{sgn}(E-V_0)}_{+ \zeta} $ and $ \Psi^{\text{sgn}(E-V_0)}_{- \zeta} $ , respectively). The dotted blue, orange, and green curves represent $G_\theta^{(1)}$, $G_\theta^{(2)}$, and $ G_\theta = G_\theta^{(1)} + G_\theta^{(2)}$, respectively, for $\theta =0 $ and in units of $ 0.25 \, e^2/(2\,\pi)$.
}
\end{figure}
%%%%%%%%%%%%%%%%%%%%%%

Using the coordinates $k_z =\frac{E}{1-\zeta}  \cos {\tilde \theta}$,
$k_y= \frac{E}{1-\zeta}  \sin  {\tilde \theta} \cos  \phi $,
$m= \frac{E}{1-\zeta} \sin  {\tilde \theta} \sin  \phi $, and
$ k_x= \pm \sqrt{k_{\text{sep}}\, (k_{\text{sep}}+2 \,m)}$,
the expression for the zero-temperature conductance per unit surface area for this case is given by
\begin{align}
G_\theta(E,V_0)  
&= \frac{e^2}{2\,\pi} \int dk_x \,dk_y \,\mathcal T(E,V_0,\theta)
= \sum_{\gamma = 1, \, 2 } G^{(\gamma)}_\theta(E,V_0)  \,,\nn 
%%%%%%%%%%%%%%%%%%%
G_\theta^{(\gamma)}(E,V_0) 
& = \frac{e^2}{2\,\pi}
\int dk_x \,dk_y\,\mathcal T_\gamma (E,V_0,\theta)
 %%%%%%%
 =  \frac{e^2}{2\,\pi}
 \int_{ \tilde \theta=0 }^{\pi/2} \int_{\phi=0}^{2\,\pi}
 d{\tilde \theta} \,d\phi\,
 \left| 
 \frac{ E ^4 \sin  (2 \tilde{\theta } ) \,\sqrt{k_{\text{sep}} 
 \left(\frac{2 \, E ^2  \sin \phi \sin  \tilde{\theta }
}
 {1-\zeta }+k_{\text{sep}}\right)}}
 {2 \,(1-\zeta ) 
 \left [
 2  \,E ^2  \sin \phi \sin  \tilde{\theta }
 +(1-\zeta ) \, k_{\text{sep}}
 \right ]}
 \right|\, \mathcal T_\gamma (E,V_0,\theta)\,.
\end{align}
%%%%%%%%%%%%%%%
Fig.~\ref{figbdir3}(a) shows the $ G_\theta $ curves as functions of $E$, with $V_0$ fixed to the value of unity, for some representative values of the remaining parameters. Fig.~\ref{figbdir3}(b) shows the density of states (DOS) for the regions on the (I) left of the junction (solid light-purple curve); and (II) right of the junction (solid light-gray and light-brown curves correspond to the eigenstates $ \Psi^{\text{sgn}(E-V_0)}_{+\zeta} $ and $ \Psi^{\text{sgn}(E-V_0)}_{-\zeta} $ , respectively).
%%%%%%%%%%%%%%%%%%%%%%%%
The DOS curves help us interpret the behaviour of the conductivity curves as explained below:
%%%%%%%%%%%%%%%%%%
\begin{itemize}

\item
For $E = 0 $, the Fermi energy on the left of the junction cuts the band-crossing point where the DOS is zero, resulting in $ G_\theta = 0 $ as well. For the regime $E<V_0$, the DOS in $z<0$ increases and the DOS in $z > 0$ decreases. As we increase $E$ from zero, each transmission probability $\mathcal T_{1} $ ($\mathcal T_{2} $) increases monotonically until the DOS for $z<0$ equals the DOS of the $\mathcal{E}^{\text{sgn}(E-V_0)}_{+\zeta}$ ($\mathcal{E}^{\text{sgn} (E-V_0) }_{-\zeta}$) band. Thus each of the two $G_\theta^{(\gamma)} $-curves reaches a local maximum --- (I) for $ G_\theta^{(1)} $ [indicated by the dotted blue curve in Fig.~\ref{figbdir3}(b) taking $\theta=0$], this happens at $ E = 0.5 $ where the solid light-purple curve intersects the light-gray curve; (II) for $G_\theta^{(2)} $ [indicated by the dotted orange curve in Fig.~\ref{figbdir3}(b) taking $\theta=0$], this happens at $ E = \left( 1-\zeta\right ) / {2} $ where the solid light-purple curve intersects the light-brown curve.
%%%%
The sum $ G_\theta = G_\theta^{(1)} + G_\theta^{(2)}$ [indicated by the dotted green curve in Fig.~\ref{figbdir3}(b) taking $\theta=0$] thus reaches a local maximum somewhere between $ E = \left( 1-\zeta\right ) / {2} $ and $ E = 0.5 $, which is what is observed in Fig.~\ref{figbdir3}(a).
%%%%%%%%%%

\item
As $E$ is cranked up further beyond this point, $G_\theta $ monotonically decreases until it reaches the local minimum value of zero at $E = V_0 $, because the DOS for $z>0$ becomes zero there. Here, unlike the RSW case, a second local maximum is not observed because the incident particle is occupying the band with the lower energy eigenvalue represented by the outer cones in the dispersion.

\item
For $E  > V_0 $, the DOS in either region increases, leading to a continuous increase of $G_\theta $.

\item The relative behaviour of the individual $G_\theta$-curves for distinct values of $ \theta$ can be understood from the fact that although the DOS for $z>0$ does not depend on $\theta $ (as we obtain it by integrating over the $k_x k_y$-plane), the momentum-space region, where the transmission coefficient $\mathcal T$ is vanishing, broadens with increasing $\theta$, resulting in a greater mismatch between the scattering states on the corresponding Fermi surface projections on the two sides of the junction [cf. Fig.~\ref{figrsw1}(b)]. The broadening / flattening of the zero-conductivity region around the minimum at $ E = V_0 $ is very sharply visible for the range of parameters considered here. Again, this effect is the result of the combination of a smaller Fermi surface projection for $z>0 $ (with decreasing $ |E-V_0|$) and a greater mismatch in the perpedicular-to-the-junction momenta (with increasing $\theta $), as explained for the case of the RSW semimetal.

\end{itemize}

%%%%%%%%%%%%%%%%%%%%%%%%%%%%%%%%%%%
\section{Junction of Dirac and Weyl semimetals}
\label{secdiracweyl}

\begin{figure}[t]
\includegraphics[width = 0.36 \textwidth]{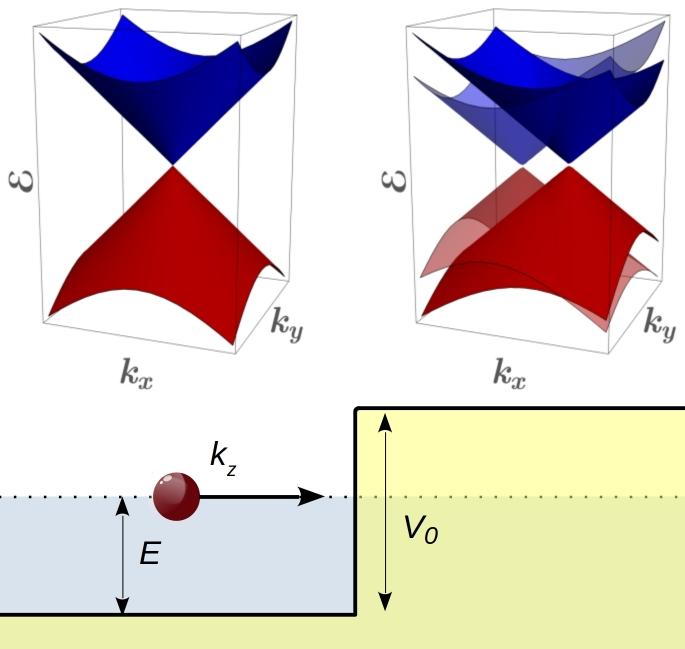}
\caption{\label{figsetup2}
Schematic diagram of the transport of a quasiparticle (red ball) into a potential step of strength $V_0$. The Fermi level (at energy $E$) is depicted by a dotted line, which cuts the conduction and valence bands on the left and right of the junction, respectively.
Inside the potential step, the twofold degenerate Dirac cone is split into a pair of Weyl cones by an intrinsic magnetization in the $xy$-plane. The energy dispersions (indicated by $\mathcal E$) of the Dirac and Weyl cones, plotted against the $k_xk_y$-plane, are also shown in the top panel.
}
\end{figure}

In this section, we consider a magnetic junction between a Dirac and
a ferromagnetic Weyl semimetal (a sandwich made of these materials was considered in Ref.~\cite{dirac-weyl}). The Dirac semimetal slab has no magnetic moment, and harbours doubly degenerate bands at a single node. On the other hand, the Weyl semimetal slab has an intrinsic magnetic moment, due to which the doubly-degenerate Dirac node is split into two nodes with positive and negative chiralities. The set-up is explained schematically in Fig.~\ref{figsetup2}.

The system is captured by the Hamiltonian
%%%%%%%%%%%%%%%%%%
\begin{align}
\label{eqdw}
\mathcal{H}_{dw} (\mathbf k) =
v \left( {\boldsymbol{\sigma}} \cdot \mathbf k \right) 
\otimes \sigma_z
+ {\mathcal J}\,\left( {\boldsymbol{\sigma}} \cdot \mathbf M \right) \otimes \sigma_0\,,
\quad \mathbf M = M(z) \left \lbrace \cos\theta, \sin\theta, 0 \right \rbrace\,, \quad
M(z)
= M_0 \,\Theta(z)\,,
\end{align}
where $v$ is the isotropic Fermi velocity, ${\mathcal J}$ is the exchange coupling constant, and $\Theta$ denotes the heaviside theta function. Aditionally, we subject the system to a rectangular scalar potential barrier $V(z) = V_0\,\Theta(z)$, which is nonzero in the region where the Weyl semimetal phase exists.
In the following, we set $v={\mathcal J}=1$ for notational simplicity.

The energy eigenvalues of the Hamiltonian $\mathcal{H}_{dw} (\mathbf k)$ are given by
\begin{align}
\mathcal{E}^\pm_\alpha
= \pm \sqrt{ \left(k_x +\alpha\, M \cos \theta \right)^2 + \left(k_z +\alpha\, M \sin \theta \right)^2 +  k_z^2 }
\, \text{ with }
%k=\sqrt{k_x^2 + k_y^2 +k_z^2} \text{ and }
\alpha =\pm\,,
\end{align}
which clearly shows that a nonzero magnetization splits the twofold degenerate Fermi surfaces (of the Dirac semimetal) in the momentum space, along the direction of the magnetization.
%%%%%%%%%%%%%%%%%%%%%%
The corresponding eigenvectors take the forms:
\begin{align}
\Psi^\pm_+ (k_x,k_y,k_z,M) &= \frac{ k_x+i \,k_y + M  e^{i\, \theta } }
{\mathcal{N}^\pm_+}
\left (
\frac{ k_z + \mathcal{E}^\pm_{+ }}
 {k_x+i\, k_y +  M\, e^{i \,\theta } }
 \quad 0 \quad 1 \quad 0
 \right)^T \nn
%%%%%%%%%%%%%
 \text{and }
\Psi^\pm_- (k_x,k_y,k_z,M) &= \frac{ k_x+i \,k_y - M  e^{i\, \theta } }
{\mathcal{N}^\pm_-}
\left ( 0 \quad 
\frac{ k_z - \mathcal{E}^\pm_{- }}
 {k_x+i\, k_y- M\, e^{i \,\theta } } \quad 0 \quad 1\right )^T ,
\end{align}
where $1/\mathcal{N}^\pm_\alpha$ represents the required normalization factor.
The nodes for the Weyl semimetal are located at $\mathbf k = M_0 \left \lbrace 
\alpha  \cos \theta, \alpha \sin \theta, 0
\right \rbrace$ in the Brillouin zone, where $\alpha = \pm $ now gives the chirality of the corresponding node.

We study the propagation of quasiparticles along the $z$-direction, incident from the leftmost slab with an energy value $E$, occupying the eigenstate $\Psi^+_-$ for $\mathcal{E}^+_-$. Hence, the scattering state is defined as follows:
\begin{align}
 & \hspace{4 cm} \Psi_{k_x,k_y,E} (z)=  \begin{cases}   
 \phi_{L} & \text{ for } z \leq 0  \\
  \phi_{R} &  \text{ for } z > 0 
\end{cases} \,,\nonumber \\
%%%%%%%%%%%%%%%%%%%%%%%%%%%%%%%%%%%%%%%%
&  \phi_{L} =   
 \Psi^+_{-}  (k_x,k_y,k_{z},0)\,e^{  i \, k_{z}  z}
 + r_+ \, \Psi^+_{+} (k_x ,k_y,- k_{z},0)\,e^{-  i \, k_{z} z}
 + r_- \, \Psi^+_{-} (k_x ,k_y,- k_{z},0)\,e^{-  i \, k_{z} z}\, ,\nonumber \\
%%%%%%%%%%%%%%%%%%%%%%%%%%%%%%%%%%%%%%%%%%%%5
&  \phi_{R}  =  
t_+ \, \Psi^{\text{sgn}(E-V_0)}_{+} ( k_x, k_y, {\tilde k}_{z}^+,M_0)
\,e^{  i \, {\tilde k}_{z}^+ z}
 + t_- \, \Psi^{\text{sgn}(E-V_0)}_{-} (k_x , k_y , {\tilde k}_{z}^-,M_0)
 \,e^{  i \, {\tilde k}_{z}^- z}\, ,\nonumber \\
%%%%%%%%%%%%%%%%%%%%%%%%%%%%%%%%% 
&  k_z  = \sqrt{ E^2-k_x^2-k_y^2} \,,\quad
%%%%%%%%%%%%%%%%%%%%%%
\tilde{k}_{z}^\alpha
= \Theta(E-V_0) \,
\sqrt{\left( E-V_0\right)^2
-k_x^2-k_y^2-M_0^2
- 2\,\alpha\, M_0  \left(  k_x \cos \theta + k_y \sin \theta \right )
}\,.
\end{align}
Here, $r_\alpha$ and $t_\alpha$ represent the reflection and transmission amplitudes, respectively, in the channels involving the $\Psi^+_\alpha $ and $\Psi^{\text{sgn}(E-V_0)}_\alpha $ eigenstates, respectively. 

%%%%%%%%%%%%%%%%%%%%%%%%%%%%%%%%%%%%%%%%%%%%%%%
%%%%%%%%%%%%%%%%

%%%%%%%%%%%%%%%%%%%fig 7 %%%%%%%%%%%%%%%%%%%%%%%%%%%%
\begin{figure}[t]
\subfigure[]{\includegraphics[width = 0.3 \textwidth]{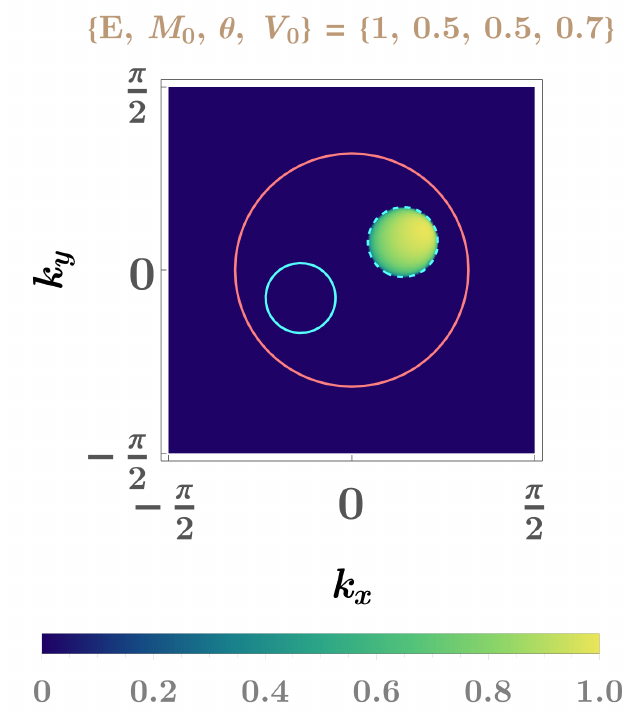}}\hspace{1.5 cm}
\subfigure[]{\includegraphics[width = 0.35 \textwidth]{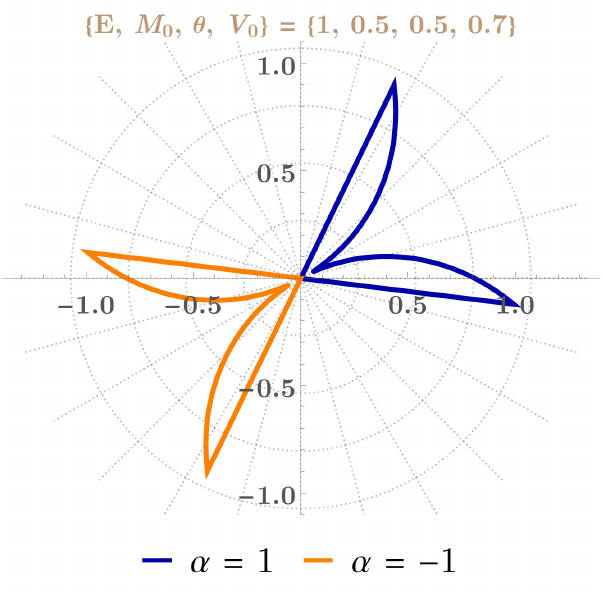}} \\
\subfigure[]{\includegraphics[width = 0.37 \textwidth]{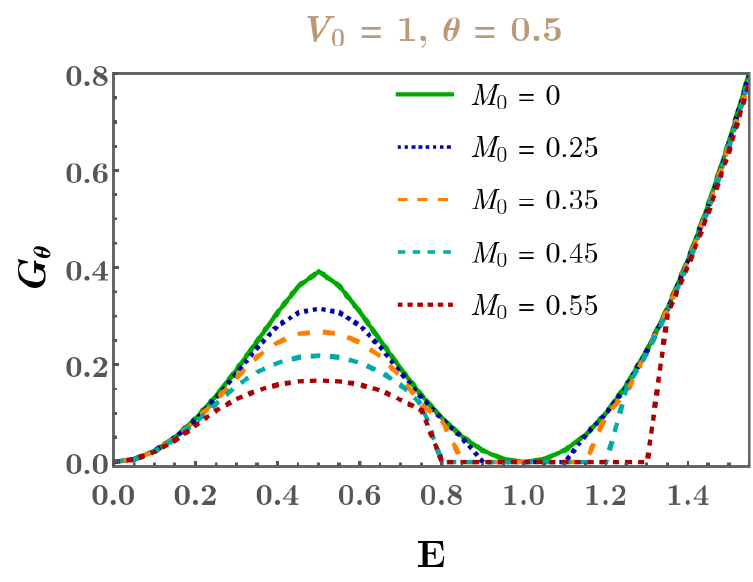}} \hspace{1.5 cm}
\subfigure[]{\includegraphics[width = 0.37 \textwidth]{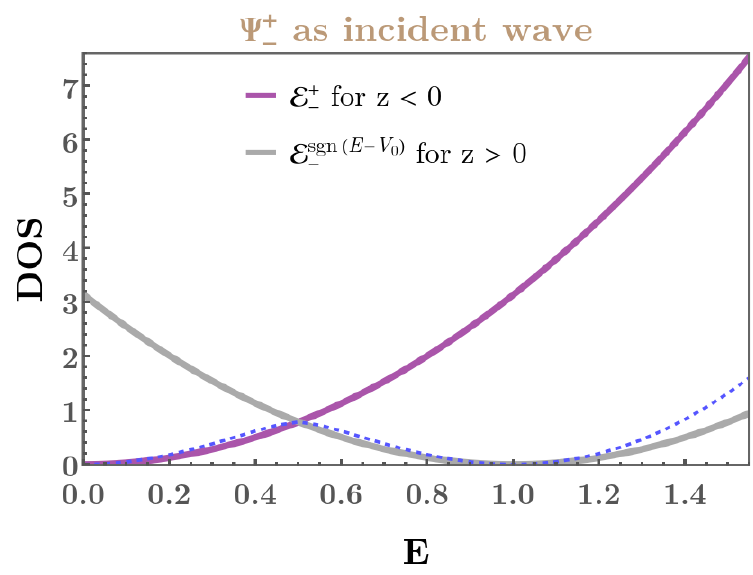}} 
\caption{\label{figdw}
Junction of Dirac and Weyl semimetals: 
\textbf{(a)} The Fermi surfaces, projected in the $k_x k_y$-plane, are shown for both sides of the junction. For the Dirac semimetal on the left of the junction, two degenerate Fermi surfaces coincide/overlap, which are plotted with the solid pink curve. On the right of the junction, there are two Fermi surface projections coming from the two Weyl nodes of opposite chiralities, which are shown by solid and dashed cyan curves (for the two chiralities). Assuming the incident quasiparticles to be occupying the $\mathcal{E}_{-}^+$ eigenstate with energy $E$, the intensity of the transmission coefficient $\mathcal T$ is also demonstrated.
%%%%%%%%%%%%%%
\textbf{(b)} Polar plots of $\sin \theta^{out}$ [cf. Eq.~\eqref{eqang3}], as functions of the incident azimuthal angle $\phi^{in}$, with the angle of incidence $\theta^{in}$ fixed at $0.5$.
%%%%%%%%%%%
\textbf{(c)} The behaviour of the zero-temperature conductance per unit surface area [in units of $e^2/(2\,\pi)$] as a function of $E$, with $V_0 =1 $, is shown for some representative values of the magnitude of magnetization of the magnetic Weyl semimetal.
%%%%%
\textbf{(d)} The curves show the DOS [in units of $(2\,\pi)^{-2}$] for the regions on the (I) left of the junction (solid light-purple curve); and (II) right of the junction (solid light-gray curve corresponding to the eigenstate $ \Psi^{\text{sgn}(E-V_0)}_{-} $) with $V_0=1 $. The dotted blue curve represents $ G_\theta $ for $\theta =0 $ in units of $ 0.5 \, e^2/(2\,\pi)$.
}
\end{figure}

Using the continuity of the wavefunction at the boundary at $z=0$, we determine the transmission and reflection amplitudes, given by the explicit expressions
\begin{align}
& r_- =
- \, \sqrt{\frac{ k_x^2+k_y^2 + \left( E +k_z\right)^2 }
{\left( k_x^2+k_y^2 + E -k_z\right) }} \,\,\,
%%%
\frac{e^{i \,\theta } \,M_0 
\left(E -k_z\right) + \left(k_x+i \,k_y\right) \left(k_z-     \tilde{k}_{z}^--V_0\right)}
{e^{i \,\theta } \,M_0 \left( E +k_z\right)-\left(k_x+i\, k_y\right)
 \left(k_z+  \tilde{k}_{z}^- + V_0\right)} 
\,, \nn 
%%%%%%%%%%%%%%%%%
& t_- =
\sqrt{\frac{
k_x^2 + k_y^2+M_0^2
-2 \, M_0 \left(  k_x  \cos \theta+  k_y \sin \theta \right)
+ \left( E - \tilde{k}_{z}^--V_0\right)^2
}
{ k_x^2+k_y^2 +\left( E -k_z\right)^2}}\,\,\,
%%%%%%%%%%%%
\frac{2 \,k_z \left(k_x+i \,k_y\right)}
{\left(k_x + i\, k_y\right) \left(k_z+ \tilde{k}_{z}^- +V_0\right)
- e^{i\, \theta } \,M_0 \left( E+k_z\right)}
 \,,\nn
%%%%%%%%%%%%%%%%
& r_+ =0\,, \quad t_+ =0 \,.
\end{align}
This tells us that there is no internode scattering, and the quasiparticle continues to propagate in the cone of the same chirality on both sides of the junction.

The transmission coefficient is obtained from
\begin{align}
\mathcal T(E,M_0,\theta, V_0)
= 1-|r_-|^2 
= \left |
\sqrt{
\frac{\tilde{k}_{z}^- / (E-V_0)}
{k_z/E}}
\, \, \,t_- \right |^2\,.
\end{align}
The behaviour of $\mathcal T$ for a chosen parameter set is shown in Fig.~\ref{figdw}(a).

Let us determine the angles of refraction using the conventions
%%%%%%%%%%%%%%%%%%%%%%%%%%%%%%%
\begin{align}
&   k_z = {E}  \cos \theta^{in}\,, \quad 
k_y = {E}  \sin \theta^{in} \,\cos \phi^{in} \,,\quad
k_x = {E}  \sin \theta^{in} \sin \phi^{in} \,,
%%%%%%%%%%%%%%%
\nn  &
b^\theta_x =k_x - \alpha\,M_0 \cos \theta \,,\quad 
b^\theta_y =k_y - \alpha\,M_0 \sin \theta \,,\quad
%%%%%%%%%%
 {\tilde k}_z =\text{sgn}(E-V_0)  \,
\sqrt{  \left( E-V_0  \right)^2-\left( b^\theta_x \right)^2
-\left( b^\theta_y \right)^2 } \,.
\end{align}
%%%%%%%%%%%%%%
Here, we have used the fact that the two Weyl nodes are located at $M_0 \left \lbrace 
\alpha  \cos \theta, \alpha \sin \theta, 0 \right \rbrace  $, with $\alpha$ indicating the chirality.
The symbols $\left \lbrace \theta^{in}, \theta^{out} \right \rbrace$ denote the angles that the velocity vectors make with the normal to the interface, and the corresponding azimuthal angles in the $xy$-plane are denoted by $\left \lbrace \phi^{in}, \phi^{out} \right \rbrace $.
The angle of refraction is therefore given by
\begin{align}
\label{eqang3}
\theta^{out} =
\cos ^{-1}\left(\frac{ {\tilde k}_z} {\left| E-V_0\right| }\right).
\end{align}
We have plotted $\sin \theta^{out}$ for the two distinct nodes in Fig.~\ref{figdw}(b), when the angle of incidence $\theta^{in} $ from the corresponding Dirac node (of the same chirality) is equal to 0.5.

Using the coordinates $k_z =\ E  \cos {\tilde \theta}$,  $k_x= E  \sin  {\tilde \theta} \cos  \phi $, and
$ k_y= E  \sin  {\tilde \theta} \sin  \phi $,
the expression for the zero-temperature conductance per unit surface area is given by
\begin{align}
G_{\theta}( E,M_0, V_0) & 
= \frac{e^2}{2\,\pi}\int dk_x \,dk_y\,\mathcal T(E,M_0,\theta, V_0)
 = \frac{e^2}{2\,\pi}
 \int_{ \tilde \theta=0 }^{\pi/2} \int_{\phi=0}^{2\,\pi}
 d{\tilde \theta} \,d\phi\,
 \left| 
 \frac{ E^2 \sin (2 {\tilde \theta} )} {2}
 \right|\,
\mathcal T(E,M_0,\theta, V_0)\,.
\end{align}
%%%%%%%%%%%%%%%%%%
Some representative curves for $G_\theta$ versus $E$ are shown in Fig.~\ref{figdw}(c), with $V_0$ fixed at the value unity.
We have also elucidated the DOS for the states with chirality $\alpha= -1$ in Fig.~\ref{figdw}(d) --- (I) solid light-purple curve representing the left of the junction; and (II) solid light-gray curve representing the right of the junction. For $E = 0 $, the Fermi energy on the left of the junction cuts the band-crossing point where the DOS is zero, resulting in $ G_\theta = 0 $.
Similar to the cases studied in the earlier section, for the regime $E<V_0$, the DOS in $z<0$ increases and the DOS in $z > 0$ decreases. As we increase $E$ from zero, the transmission probability $\mathcal T $) thus increases monotonically until the DOS for $z<0$ equals the DOS of the $\mathcal{E}^{\text{sgn}(E-V_0)}_{-}$ band, causing the $G_\theta $-curve [indicated by the dotted blue curve in Fig.~\ref{figdw}(d) setting $\theta=0$] to reach a local maximum at $ E = 0.5 $, where the solid light-purple curve intersects the light-gray curve.
%%%%
As $E$ increases further beyond this point, $G_\theta $ monotonically decreases until it reaches the local minimum value of zero at $E = V_0 $, because the DOS for $z>0$ becomes zero there. In the regime $E  > V_0 $, the DOS in either region increases, leading to a continuous increase of $G_\theta $.
%%%%%%
The relative behaviour of the individual $G_\theta$-curves for different values of $ M_0 $ and $\theta $ can be understood from the fact that although the DOS for $z>0$ does not depend on these parameters (as we obtain it by integrating over the $k_x k_y$-plane), the momentum-space region, where the transmission coefficient $\mathcal T$ is vanishing, broadens with increasing $\theta$, resulting in a greater mismatch between the momenta of the scattering states on the corresponding Fermi surface projections on the two sides of the junction [cf. Fig.~\ref{figdw}(a)]. In this case, the mismatch in the momentum vectors of the incoming and outgoing states can be quantified by the vector $M_0 \left \lbrace \cos\theta, \sin\theta \right \rbrace $ in the $k_x k_y$-plane. Therefore, a larger value of $M_0 $ and /or $\theta $ (within the range $\left [ 0, \, \pi/2 \right ] $) leads to a broadening of the zero-conductivity regions on both sides of the local minimum at $ E=V_0$.

%%%%%%%%%%%%%%%%%%%%%%%%%%%%%%%%%%%%%%%%%%%%%%%%%%%%%%%%%%%%%%%%%%%%%%%%%%%%%%%%
\section{Sandwich of Weyl/multi-Weyl semimetals}
\label{secmultiweyl}

The Weyl semimetal ($J=1$) Hamiltonian at the node with positive chirality is given by
\begin{equation}
\label{a1}
\mathcal{H}_1(\mathbf{k})=  v\, \mathbf{k} \cdot \boldsymbol{\sigma} ,
\end{equation}
%%%%%%%%%%%%%%%5
where $v$ is the isotropic Fermi velocity. 
The energy eigenvalues are given by
\begin{equation}
\label{a2}
\mathcal{E}_1^s(\mathbf{k})= \pm \, v\, |\mathbf k| \text{ with } s=\pm\,,
\end{equation}
%%%%%%%%%%%%%%%%5
where the ``$+$'' and ``$-$'' signs correspond to the conduction and valence bands, respectively. A set of normalized eigenvectors corresponding to $\mathcal{E}_1^\pm(\mathbf k)$ are given by
%%%%%%%%%%%%%%%%%%55
\begin{align}
\label{a3}
\psi_{1}^s (\mathbf{k})  
=\frac{1}{\mathcal{N}_1^s }
\left( 
\frac{v^{-1}\, \mathcal{E}^s_1+ k_z  }
{  k_x +  i  \,k_y }
\quad  1 \right )^T ,
\end{align}
where $1/\mathcal{N}_1^s$ denotes the normalization factor.

%%%%%%%%%%%%%%%%%%%%%%%%%%%%%%%%%%%%%%%%%%%%%%%
\begin{figure}[t]
\includegraphics[width = 0.75 \textwidth]{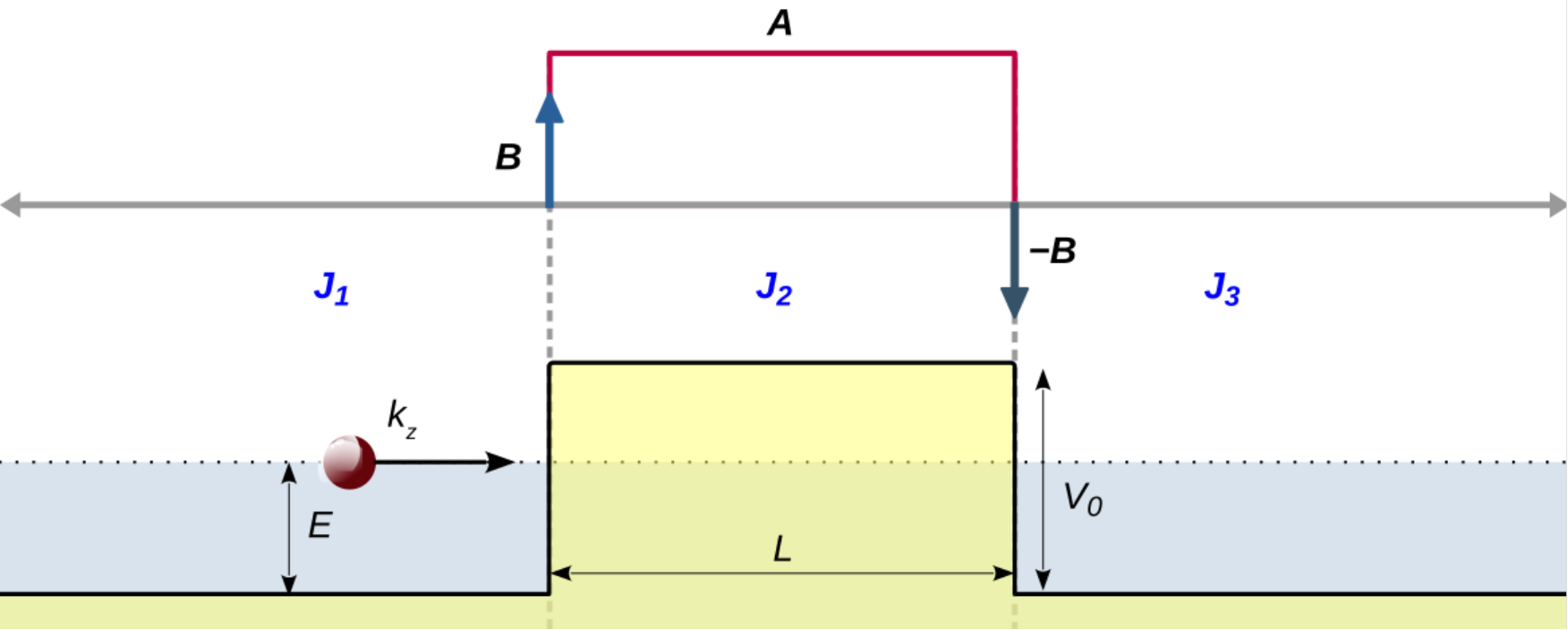}
\caption{\label{figsetup3}
Schematic diagram of the transport of a quasiparticle (red ball) across a scalar potential barrier of strength $V_0$, with a constant vector potential $\mathbf{A}$ superposed in the same region. The Fermi level (at energy $E$) is depicted by a dotted line, which cuts the conduction(valence) band outside(inside) the barrier. The $J$-values in the three regions are denoted by $\lbrace J_1, J_2, J_3 \rbrace$. Theoretically, the required vector potential can be created by applying equal and opposite delta function magnetic fields ($\mathbf{B}$ and $-\mathbf{B}$) at the edges of the barrier region, oriented perpendicular to the direction of transmission.
}
\end{figure}
%%%%%%%%%%%%%%%%

The multi-Weyl semimetals are generalizations of the Weyl Hamiltonian to nodes with higher values of topological charges, resulting from a nonlinear dispersion in the plane perpendicular to the $z$-axis.
The effective continuum Hamiltonian for an isolated multi-Weyl
node of chirality $\chi =\pm 1$ and topological charge $J$ is given by
\cite{PhysRevB.95.201102,PhysRevB.97.045150}
\begin{align}
\label{eqhammulti}
\mathcal{H}_J(\mathbf{k})= \frac{ v_{\perp}
\left [  \left (k_x -  i \, k_y \right)^J 
\left( \sigma_x + i \, \sigma_y \right)
+ \left (k_x +  i \, k_y \right)^J 
\left( \sigma_x - i \, \sigma_y \right) \right ]}
{2\,k_0^{J-1}}    
+ \chi\, v_z\,  k_z\, \sigma_z \,,
\end{align}
where $v_z$ and $v_{\perp}$ are the Fermi velocities in the $z$-direction and $xy$-plane, respectively, and $k_0$ is a material-specific parameter with the dimension of momentum.
$\mathcal{H}_1(\mathbf{k})$ can be obtained from $\mathcal{H}_J(\mathbf{k})$ by setting $v_\perp=v_z =v $.
For the sake of completeness, the explicit forms are:
\begin{align}
&    \mathcal{H}_2(\mathbf{k})= \frac{ v_{\perp} 
\left [  \left (k_y^2-k_x^2 \right)\sigma_x+ 2\, k_x\, k_y\, \sigma_y \right ]}
{k_0}    
+ \chi\, v_z\,  k_z\, \sigma_z \,,\nn
%%%%%%
&
 \mathcal{H}_3(\mathbf{k})=
\frac{ v_{\perp}\,  \left [ \left (k_x^3\, \sigma_x - k_y^3\, \sigma_y \right)
+ 3 \left (  k_x\, \sigma_y- k_y\, \sigma_x \right ) k_x\, k_y \right ]}
{k_0^2}
+ \chi\, v_z\,   k_z\, \sigma_z \,.
\end{align}

First let us focus on the $\chi = 1$ case.
The eigenvalues are then given by
%%%%%%%%%%%%
\begin{align}
\mathcal{E}^s_J(\mathbf{k})=  s\, 
\sqrt{  \frac{v_{\perp}^2\,k_\perp^{2J}} {k_0^{2J-2}} + v_z^2\, k_z^2}\,,
\text{ where } k_\perp =\sqrt{k_x^2 +k_y^2}
\text{ and } s=\pm.
\end{align}
The corresponding normalized eigenvectors take the form
\begin{align}
\label{psij}
\psi_{J}^s(k_x, k_y, k_z) = \frac{1}{\mathcal{N}^s_J }
\left(
\frac{k_0^{J-1} 
\left ( \mathcal{E}^s_J+ v_z \,k_z \right )}
{ v_\perp \left(k_x +  i  \,k_y \right )^{J}
} 
\quad  1 \right)^T ,
\end{align}
where $1/\mathcal{N}_J^s$ denotes the normalization factor.
The labels $\pm$ denote the conduction and valence bands respectively. 

%%%%%%%%%%%%%%%%%%%%%%%%%%%%%%%%%%%%%%%%%%%%%%%
\begin{figure}[t]
\includegraphics[width = 0.75 \textwidth]{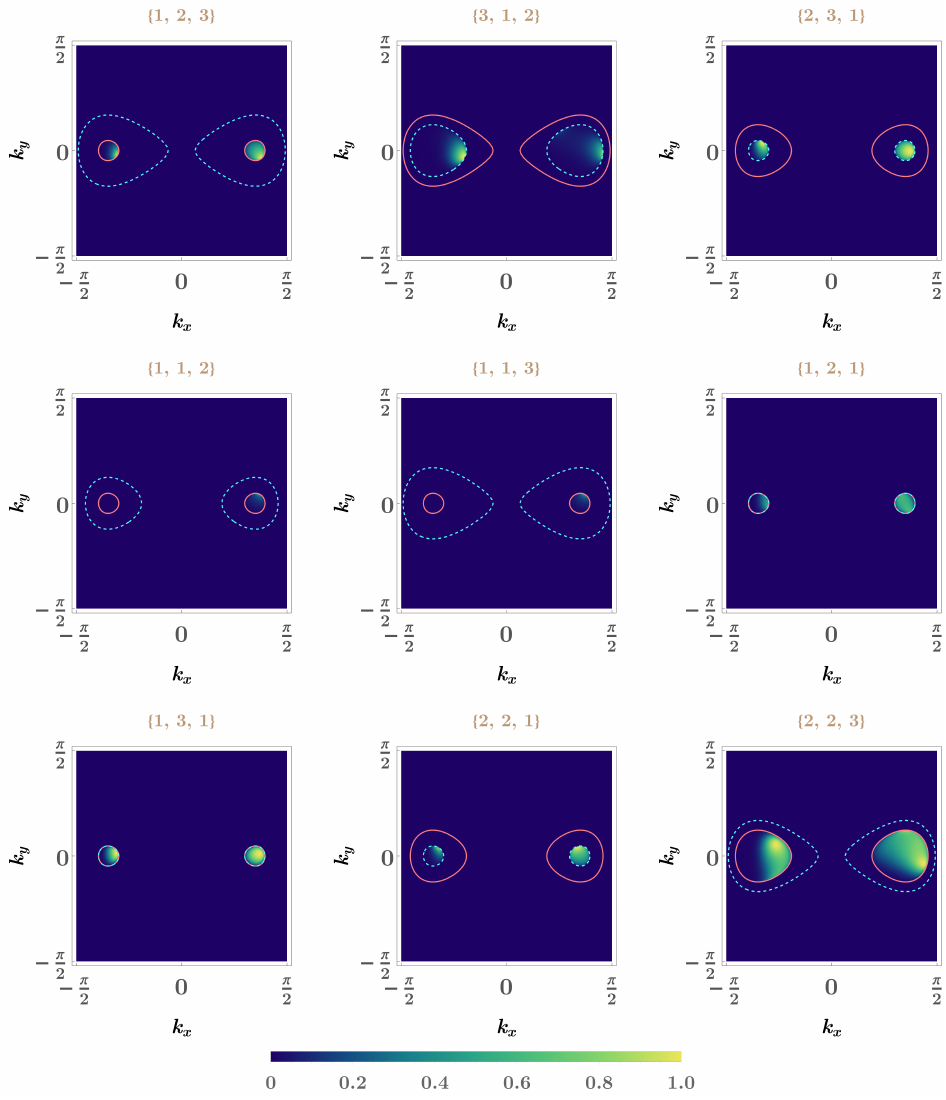}
\caption{\label{figmwsm1}
Sandwiches of Weyl/multi-Weyl semimetals with $k_{\text{sep}}=1.1$, $ V_0 = 0.25 $, $L= 25$, $a_x = 0.5$, $a_y =0.1 $, and $E=0.15$:
The Fermi surface projections in the $k_x k_y$-plane, for the incidence (solid pink curves) and transmission (dashed cyan curves) regions of the sandwich configurations.
The $J$-values in the three regions are denoted by $\lbrace J_1, J_2, J_3 \rbrace$, which have been indicated as plot-labels in all the panels. Assuming the incident quasiparticles to be occupying the $\mathcal{E}_{J_1}^+$ eigenstate with energy $E$, the spectrum of the transmission coefficient $\mathcal T$ is also shown.
}
\end{figure}
%%%%%%%%%%%%%%%%%%%%%%

We now consider a sandwich configuration of Weyl/multi-Weyl semimetals consisting of three different $J$-values, denoted by $\lbrace J_1, J_2, J_3 \rbrace$, with the middle region (with $J=J_3$)
being subjected to both electric and vector potentials (see Fig.~\ref{figsetup3}). The quasiparticle (with energy $E$) from a conduction band is incident from the leftmost $J=J_1$-region, located at $z<0$, propagating in the $z$-direction. Between $z=0$ and $z=L$, it enters the region with $J=J_3$,
where a potential barrier exists, such that
\begin{equation}
\label{eqpot}
V(z)
=\begin{cases}V_0 &0\leq z\leq L \\0 & \text{otherwise}
\end{cases}\,.
\end{equation}
Thus, the momentum components $k_x$ and $k_y$ are conserved across the barrier.
This is supplemented by equal and opposite magnetic fields localized at the edges of the rectangular electric potential, and directed perpendicular to the $z$-axis \cite{mansoor,kai}. The magnetic fields can be theoretically modelled as Dirac delta functions of opposite signs at $z=0$ and $z=L$, respectively. A choice of the vector potential can be made as
\begin{align}
\mathbf {A}( z ) \equiv  \lbrace a_x, a_y,0  \rbrace = \begin{cases} 
\lbrace B_y, -B_x, 0  \rbrace &  \text{ for } 0 \leq z \leq L \\
\mathbf 0 & \text{ otherwise} 
\end{cases} ,
\label{eqvecpot}
\end{align}
which clearly gives rise to the desired the magnetic field configuration of the form $\mathbf{B}  = 
\left(  B_x\, \hat{\mathbf i}  + B_y\, \hat{\mathbf j}   \right )
\left [  \delta\left(z\right) - \delta\left( z-L \right) \right ] $.
Using the Peierls substitution, the effect of the vector potential is incorporated by replacing the transverse momenta as $k_x \rightarrow k_x- a_x,$ and $ k_y\rightarrow k_y- a_y$ (setting the electric charge $e=1$). Hence, the resulting effective Hamiltonian within the barrier region is given by
$\mathcal{H}_J(  k_x- a_x, k_y - a_y, -i\,\partial_z) +V_0 $.
Finally, the quasiparticle leaves the barrier region at $z> L$, and enters the rightmost region with $J=J_3$.
We compute the transmission coefficient for this propagation from left to right.

Similar to the RSW and birefringent semimetal cases of Sec.~\ref{secrswbdirac}, we consider a more generic system with two cones of opposite chiralities in each region. This is implemented in a straightforward way by replacing $k_x$ by $m$ in the Hamiltonian defined in Eq.~\eqref{eqhammulti}.
%%%%%%%%%%%%%%%%%%%%%%%%%%%%%%%
A scattering state $ \Psi_{k_x,k_y,E} (z)$ can now be constructed as follows:
\begin{align}
&  \hspace{2 cm} \Psi_{k_x,k_y,E} (z)=  \begin{cases}   
 \phi_{J_1,L} & \text{ for } z <0  \\
 \phi_{J_2,M} & \text{ for } 0 \leq z \leq L \\
  \phi_{J_3,R} &  \text{ for } z > L 
\end{cases} \,, 
\end{align}
\begin{align}
%%%%%%%%%%%%%%%%%%%%%%%%%%%%%%%%%%%%%%%%
&  \phi_{J_1,L} = \frac{   
 \psi_{J_1}^+ (m,k_y,k_z)\,e^{  i \, k_z z}
 +r \, \psi_{J_1}^+ (m,k_y,-k_z)\,e^{-  i \, k_z z}}
{\sqrt{  { \mathcal{V} }_{J_1}(m, k_y,k_z ) }}\, ,\nn &
%%%%%%%%%%%%%%%%%%%%%%%%%%%%%%%%%%%%%%%%%%%%5
  \phi_{J_2,M}  =   
 \lambda \,\psi_{J_2}^{\text{sgn}(E-V_0)} (\tilde m,\tilde k_y,\tilde{k}_z) \,
 e^{  i \,\tilde k_z  z } 
 + 
 \gamma \, \psi_{J_2}^{\text{sgn}(E-V_0)} (\tilde m,\tilde k_y,-\tilde{k}_z) \,
 e^{-  i \,\tilde k_z z } 
 , \nn &
%%%%
\phi_{J_3,R} = \frac{  t \,\psi_{J_3}^+ ( m,k_y,k_{\ell}) 
 } 
{\sqrt{  { \mathcal{V} }_{J_3}  (m,k_y,k_\ell)}}
\, e^{  i \, k_\ell \left( z -L\right)}\,, \nn
%%%%%%%%%%%%%%%%%%%%%%%%%%%%%%%%% 
& m= \frac{k_x^2 - k_{\text{sep}}^2}{2\,k_{\text{sep}}}\,, \quad
k_z  = \frac{1} {v_z}\,
 {\sqrt{E^2 -  \frac{v_{\perp}^2 \left( m^2+k_y^2 \right)^{J_1}} 
 {k_0^{ 2\, J_1-2}}}} \,, \quad
%%%%%
k_\ell = \frac{1} {v_z}\,
 {\sqrt{E^2 -  \frac{v_{\perp}^2 \left( m^2+k_y^2 \right)^{J_3}} 
 {k_0^{ 2\, J_3-2}}}} \,,  \quad
 \tilde k_x =  k_x -  a_x\,,\quad 
\tilde k_y = k_y -  a_y\,,
 \nn
%%%%%%%%%%%%%%%%%%%%%%%%%%%%55
& \tilde m =  \frac{\tilde k_x^2 - k_{\text{sep}}^2}{2\,k_{\text{sep}}}\,,\quad
\tilde k_z =\frac{1} {v_z} \,
\sqrt{ \left(E- V_0 \right) ^2 
-  \frac{v_{\perp}^2\left( \tilde m^2 + \tilde k_y^2 \right)^{J_2}} 
{k_0^{ 2\, J_2-2} }}\,,
\quad
%%%%%%%%%%%%% 
{ \mathcal{V} }_J  (m ,k_y, k_{z}^{\text{var}}) =   
\big |\partial_{k_{z}^{\text{var}}}  \mathcal{E}_{J}^+ (m,k_y,k_{z}^{\text{var}})\big|
\,,
\end{align}
where we have used the velocity $ { \mathcal{V} }_J (k_x, k_y,k_{z}^{\text{var}}) $ to normalize the incident, reflected, and transmitted plane waves. Here, $r $ and $t$ are the amplitudes of the reflected and transmitted waves, respectively.
The four unknown coefficients $\left \lbrace r, \,t, \,\lambda,\, \gamma \right \rbrace $ can be computed by solving the four equations, which are obtained from the continuity conditions for the two components of the wavefunction at the two boundaries (viz., $z=0$ and $z=L$). In the following, we will set $v=v_z=v_\perp =k_0= 1$ for the sake of simplicity.

%%%%%%%%%%%%%%%%%%%%%%%%%%%%%%%%%%%%%%%%%%%%%%%
\begin{figure}[t]
\includegraphics[width = 0.9 \textwidth]{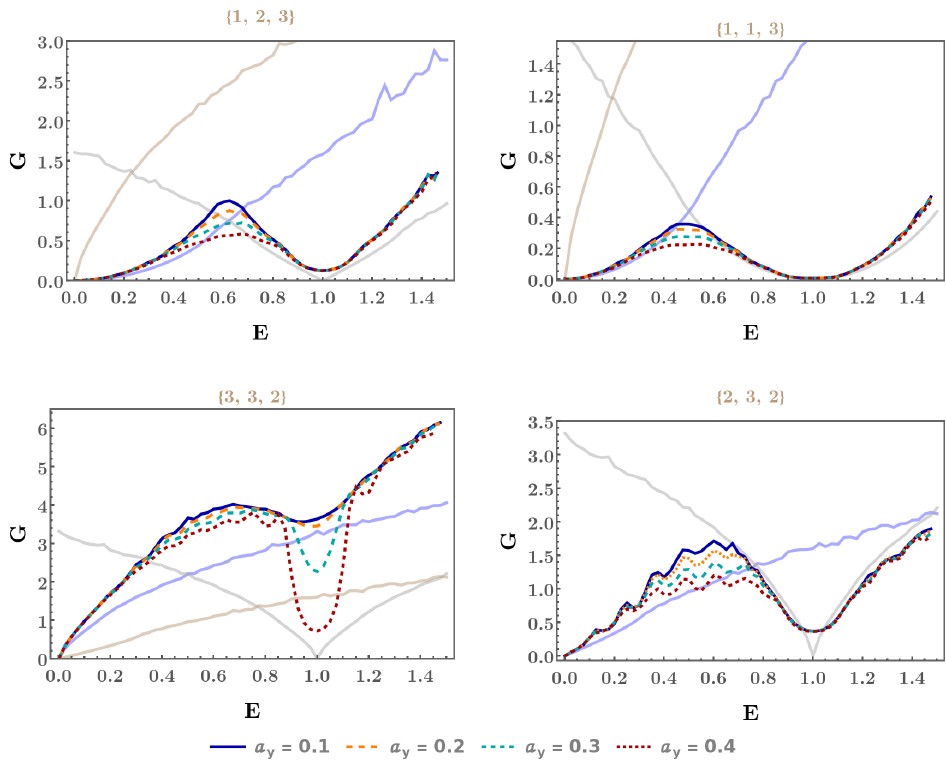}
\caption{\label{figmwsm2}
Sandwiches of Weyl/multi-Weyl semimetals with $k_{\text{sep}}=1.1$, $ V_0 = 1 $, $L= 25$, and $a_x = 0.1$: The curves show the behaviour of the zero-temperature conductance per unit surface area [in units of $e^2/(2\,\pi)$] versus $ E $ for different values of $a_y$ (as indicated in the plotlegends). The $J$-values in the three regions are denoted by $\lbrace J_1, J_2, J_3 \rbrace$, which appear as plot-labels. For each subfigure, the light-blue, light-gray, and light-brown curves in the background indicate the densities of states [in units of $1/(8\,\pi^2)$] in the regions $z<0$, $ 0<z<L $, and $z>L $, respectively. If $J_1 = J_3 $, the light-blue curve represents the densities of states in both the regions $z<0$ and $z> L $. 
}
\end{figure}
%%%%%%%%%%%%%%%%

%%%%%%%%%%%%%%%%%%%%%%%%%%%%%%%%%%%%%%%%%%%%%%%
\begin{figure}[t]
\includegraphics[width = 0.9 \textwidth]{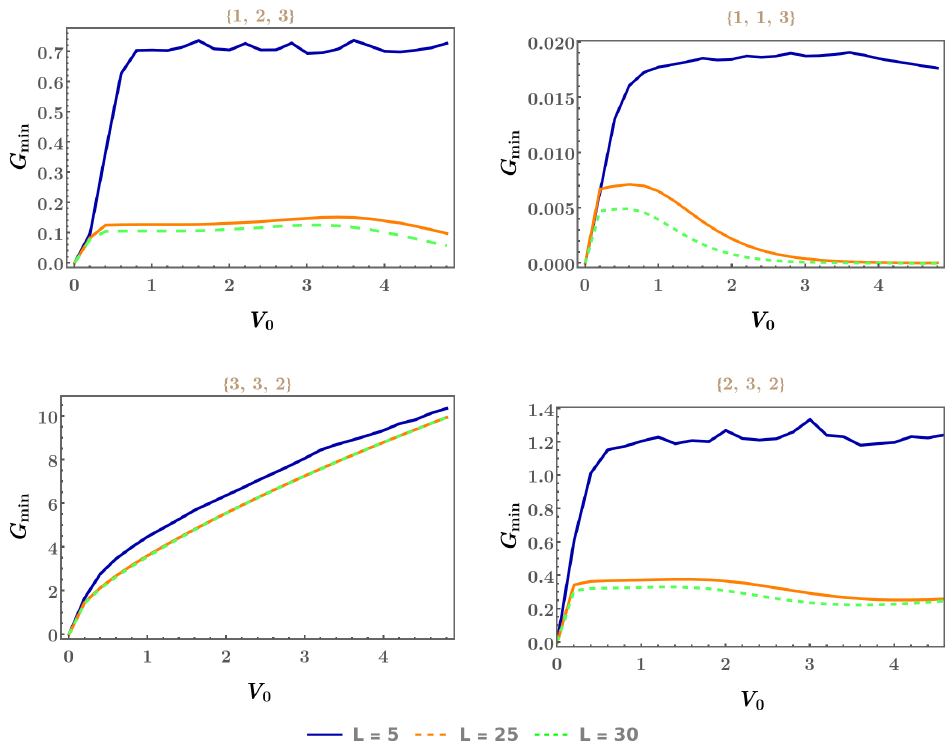}
\caption{\label{figmwsm3}
Sandwiches of Weyl/multi-Weyl semimetals with $k_{\text{sep}}=1.1$, $ V_0 = 1 $, $L= 25$, and $ a_x = a_y = 0 $:
The curves show the variation of $G_{\text{min}}$ [in units of $e^2/(2\,\pi)$] with $ V_0 $ for different values of $L$ (as shown in the plotlegends). The $J$-values in the three regions are denoted by $\lbrace J_1, J_2, J_3 \rbrace$, which have been indicated as plot-labels.
}
\end{figure}
%%%%%%%%%%%%%%%%

The behaviour of the transmission coefficient $\mathcal T(E,V_0,\mathbf A) = |t|^2=1-|r|^2$, as a function of $V_0$, is shown in Fig.~\ref{figmwsm1} for some specific sandwich configurations. We find that $\mathcal T$ is restricted to the overlapping area enclosed by the Fermi surface projections in the incidence and transmission regions.
%%%%%%%
Using the coordinates $k_z = E  \cos {\tilde \theta}$,
$k_x= \left(E \sin \tilde{\theta }\right)^{{1}/{J_1}} 
\cos\phi $, and
$ k_y= \left(E \sin \tilde{\theta }\right)^{{1}/{J_1}} 
\sin \phi $, the expression for the zero-temperature conductance per unit surface area for this case is given by
\begin{align}
G(E,V_0,\mathbf A) & 
= \frac{e^2}{2\,\pi} \int dk_x \,dk_y\,\mathcal T(E,V_0,\mathbf A)
 = \frac{e^2}{2\,\pi}
 \int_{ \tilde \theta=0 }^{\pi/2} \int_{\phi=0}^{2\,\pi}
 d{\tilde \theta} \,d\phi\,
 \left| 
\frac{ \sqrt{k_{\text{sep}}} \, \cot  \tilde{\theta }  
\left( E \sin  \tilde{\theta } \right)^{\frac{2}{J_1 }}}
{ J_1 \,\sqrt{ 
2 \cos \phi  
\left( E \sin  \tilde{\theta } \right)^{\frac{1}{J_1 }}
+k_{\text{sep}} }
}
 \right|\,
\mathcal T(E,V_0,\mathbf A)\,.
\end{align}
%%%%
Fig.~\ref{figmwsm2} shows the $G$ curves as functions of $ E $ for $V_0= 1 $, $L=25 $, $a_x= 0.1 $, and $ a_y = \lbrace  0.1, \, 0.2, \, 0.3,\,0.4 \rbrace $. The light-blue, light-gray, and light-brown curves in the background indicate the densities of states in the regions $z<0$, $ 0<z<L $, and $z>L $, respectively. The energy-dependence of the DOS goes as $ \sim E^{2/J_1}$, $ \sim |E-V_0|^{2/J_2}$, and $ \sim E^{2/J_3}$ in these three regions.
If $J_1 = J_3 $, the light-blue curve represents the densities of states in both the regions $z<0$ and $z> L $. Here, although the DOS information provides some amount of intuition regarding the behavior of $G$, it cannot tell us the full story as the global nature of $G$ depends very strongly on the values of $V_0$, $ L $, and $J_3$ \cite{krish-sandwich,Deng2020,ips-aritra}. In particular, let us discuss the factors affecting the value of $G$ at $ E= V_0$ (labelling it as $ G_{\text{min}} $), where the DOS in the barrier region goes to zero: 
\\(I) As $ V_0 $ is increased from zero, $ G_{\text{min}} $ also increases from zero. But after $V_0$ reaches a cut-off, the $ G_{\text{min}} $ shows a nonmotonic behaviour [cf. Fig.~\ref{figmwsm3}].
\\(II) The higher the value of $L$, the more pronounced (and closer to zero) is the local minimum, as seen from Fig.~\ref{figmwsm3}.
\\(II) Despite the vanishing DOS in the $0<z<L$ region, a nonzero $G$ at $ E= V_0$ is caused by transport via evanescent waves (exponentially decaying or diverging with $z$) in the barrier region --- greater the value of $J_3$, the more transparent the barrier becomes (i.e., larger values of $ G_{\text{min}} $), which we can clearly see by comparing the values of $J_3$ in Figs.~\ref{figmwsm2} and Fig.~\ref{figmwsm3}.
Finally, the relative behaviour of the $G$-curves for different values of $ \lbrace a_x, \, a_y \rbrace $ can be understood from the fact that although the DOS for $z>0$ does not depend on these parameters (as we obtain it by integrating over the $k_x k_y$-plane), a mismatch between the momenta of the scattering states inside and outside the barrier [cf. Fig.~\ref{figmwsm1}] is quantified by the vector $ \lbrace a_x, \, a_y \rbrace $.
Hence, larger magnitudes of $a_x$ and/or $a_y$ lead to an overall suppression of the conductance, as the momentum region where the transmission coefficient $\mathcal T$ vanishes enlarges. This is seen in Fig.~\ref{figmwsm2}.

%%%%%%%%%%%%%%%%%%%%
\section{Summary and outlook}
\label{secsum}

In this paper, we have computed the transmission coefficients of quasiparticles in various geometries involving junctions of three-dimensional semimetals. We have also investigated the behaviour of the zero-temperature conductance across the junctions. Our results show that a nonzero transmission spectrum exists only where the areas, bounded by the Fermi surface projections (in the plane perpendicular to the propagation axis) of the incidence and transmission regions, overlap. A possible application of these findings is that the existence of stable and tunable junctions has the potential to help engineer unidirectional carrier propagation, topologically protected against impurity backscattering \cite{expt1}, because of the chiral nature of the charge carriers. 

For future explorations, we enumerate some promising directions below:
\begin{enumerate}[{(1)}]

\item A generalization of the RSW semimetal studied in this paper can be found in Ref.~\cite{isobe-fu}, where the full rotational symmetry of RSW is broken to the O$_h$ symmetry. The dispersion features anisotropic velocity parameters. A junction set-up with such an anisotropic system, similar to what we have considered in Sec.~\ref{secrswbdirac}, will show a richer structure in the transport features.

\item For the Dirac-Weyl junction studied in Sec.~\ref{secdiracweyl}, one can look at the transport properties when the junction is encountered while propagating along a direction lying in the plane of magnetization of the Weyl semimetal part of the system. In other words, the quasiparticle travels along a direction within the plane of magnization (i.e., the $xy$-plane in the coordinate system chosen in this paper).

\item A Luttinger semimetal harbours doubly-degenerate bands, similar to the Dirac semimetals, but with anisotropic nonlinear dispersions \cite{lutt1,lutt2}. By breaking time-reversal or cubic symmetries of this system, one can lift the degeneracy to get a pair of double-Weyl nodes of opposite chiralities, separated in momentum space \cite{MoonXuKimBalents}. This scenario is similar to breaking the doubly degenerate Dirac node into a pair of Weyl nodes via nonzero magnetization, and hence we can construct a set-up similar to what we have studied in Sec.~\ref{secdiracweyl}.
Due to anisotropy and nonlinearity of the band dispersions, this system is expected to show exotic transport characteristics.

\item For the Weyl/multi-Weyl semimetal sandwiches considered in Sec.~\ref{secmultiweyl}, one can compute the transport features when the junctions are encountered while propagating along a direction with nonlinear dispersion
(i.e., embedded within the $xy$-plane when we use the coordinate system chosen in this paper). This will be a more challenging computation, in the same spirit as has been carried out in Refs.~\cite{Deng2020,ips-aritra}.

\end{enumerate}

Other directions to explore these transport properties in the presence of disorder and/or many-body interactions \cite{ips-seb,ips-biref,ips-klaus,rahul-sid,ipsita-rahul,ips-qbt-sc}. Yet another possible direction is to apply time-periodic drives \cite{ips-sandip,ips-sandip-sajid,shivam-serena}, generalizing the static scalar potentials that we have considered in this paper.

%%%%%%%%%%%%%%%%%%%%%%%%appendix%%%%%%%%%%%%%%%%%5

\appendix

\def\theequation{A\arabic{equation}}

\section*{Appendix: Boundary conditions and continuity of probability flux}
\label{app1}

Let us consider the low-energy expansion of the Hamiltonian around a nodal point, which can be described as 
\begin{align} 
	\label{eq:kdots}                                  
	\mathcal{H} (\mathbf k ) =    \mathbf{d} (\mathbf k) \cdot \boldsymbol{\mathcal{S}}\,, 
\end{align}
where $\mathbf{d} (\mathbf k) =\lbrace d_x(\mathbf k), \, d_y(\mathbf k), \, d_z(\mathbf k_z) \rbrace$ represents three functions of the momenta, and $\boldsymbol{\mathcal{S}}$ is a three-vector, made up of the matrices $S_j$ obeying the algebra $\left [S_j, \,S_k \right ]=i\,\epsilon_{jkl} \,S_l$ [with $j, k, l \in \lbrace x,\, y,\, z \rbrace $], that represents the pseudospin degrees of freedom. Here, we have dealt with the following cases:
\\(1) For the RSW semimetal, $\boldsymbol{\mathcal{S}}$ is given by the $4\times4$ pseudospin-3/2 matrices $\mathbf J =\lbrace J_x , \, J_y,\, J_y \rbrace $ shown in Eq.~\eqref{eqrswxJ}.
\\(2) For the pseudospin-1/2 Weyl/multi-Weyl quasiparticles, $ \boldsymbol{\mathcal{S}} =  \boldsymbol{\sigma} $, where $\boldsymbol{\sigma} = \lbrace \sigma_x , \, \sigma_y ,\, \sigma_y \rbrace $ represents the three $2\times2$ Pauli matrices.
The above form for $ \boldsymbol{\mathcal{S}} $ is clear when we note that for the multi-Weyl case, the Hamiltonian around a single node can be written as 
$$\mathcal{H} ( \mathbf k) = \frac{v_\perp \, k_\perp^J  }  {k_0^{J-1 }} 
\left [\cos (J \,\phi_k ) \, \sigma_x 
+ \sin (J\, \phi_k) \, \sigma_y \right ] + \chi \,v_z \, k_z \, \sigma_z \,,$$
where $\phi_k = \arctan(k_y,k_x)$ and $k_\perp = \sqrt{k_x^2 + k_y^2 }\,$.
\\(3) For the birefringent and Dirac semimetals, we have $\boldsymbol{\mathcal{S}}\propto \boldsymbol{\sigma} \otimes  \boldsymbol{\sigma} $, which appears as a tensor product of the Pauli matrices in two different pseudospin-1/2 spaces.

Now for the cases considered in this manuscript, we have $d_z (\mathbf k) = v_z \, k_z $ (where $v_z$ is the Fermi velocity along the $z$-direction), and the quasiparticles are moving along the $z$-axis when they encounter junctions (which are aligned perpendicular to the $z$-axis). In general, the potential $V(z)$ varies across the junction, which we take to be a finite value on either side. To capture this situation, we need to consider the Hamiltonian with the $z$-part written in position space as follows:
\begin{align}
\tilde {\mathcal{H}} &=
 \mathcal{H} ( k_x, k_y, k_z \rightarrow -i \,\partial_z  ) + V(z) \,,\quad
V(z) = \begin{cases}
V_1 & \text{ for } z \leq 0 \\
V_2 & \text{ for } z > 0 
\end{cases}\,.
\end{align}  
Then the eigenvalue equation is $ \left( \tilde {\mathcal{H}} -E \right )
\tilde \Psi_E (k_x, k_y, z) = 0 $, where $\tilde \Psi_E $ is an eigenfunction of $\tilde {\mathcal{H}}$ with eigenvalue $E$. It is a first order differential equation in $z$ and we intend to find solutions which are finite, continuous, and differentible over the entire interval $ z \in(-\infty, \infty)$ \cite{messiah}. In other words, all solutions of a differential equation must be differentiable implying that continuity is a requirement for differentiability.
%%%%%%%%%%%%%
Compare this when the Hamiltonian has a second derivative (e.g., describing a Schrodinger particle) and the differential equation reduces to the problem of the Sturm–Liouville type \cite{messiah}. For a junction located at $z=0$, we integrate the above equation over an infinitesimal interval $2\,\epsilon $ about $z =0 $, and subsequently take the limit $\epsilon \rightarrow 0 $ as follows:
\begin{align}
%%%%%%%%%%%%%%%%%%%%%%%%%%%%%
& \lim_{\epsilon \rightarrow 0} \int_{-\epsilon }^\epsilon
 dz \left[ d_x \,{\mathcal{S}}_x + d_y \,{\mathcal{S}}_y + v_{z} \,{\mathcal{S}}_z\, \partial_{z}  + V(z)
- E  \right ]
 \tilde \Psi_E (k_x,k_y, z) = 0 \,,
\end{align}
This leads to the boundary conditions 
\begin{align}
\label{eqbccond}
v_{z_1}  \,
 \tilde \Psi_E^{(1)} (k_x,k_y, z) \big \vert_{z = 0^{-}}
 = v_{z_2}  \,
 \tilde \Psi_E^{(2)} (k_x,k_y, z) \big \vert_{z = 0^{+}} \,,
\end{align}
where the subscripts and the superscripts $a \in \lbrace 1, \, 2 \rbrace $ refer to the materials on the left and right of the junction, respectively. If we have a $2 N$-component wavefunction, we get $2N$ equations.
Note that unlike the Schrödinger equation case, we do not have any more boundary conditions. In particular, we do not have a condition for the continuity of the derivative of the wavefunction at the boundary (and this scenario is similar to the case of the relativistic Dirac equation case). This is a fundamental equation and, naturally, holds for all cases discussed in the paper.

Let us assume that $ \left[ \tilde {\mathcal H}^{(a)}  - V(z) \right ]$ represents a $2N$-band system with the energy eigenvalues given by
\begin{align}
\varepsilon_{a,n}^s (k_x, k_y, k_{z_a})  = s\,c_{a,n} \, \sqrt{ \left( d_x^{(a)} \right)^2 +  \left( d_y^{(a)} \right)^2
+ v_{z_a}^2 \, k_{z_a}^2 }  
\text{ for }  n \in [1, N] \text{ and } s=\pm\,,
\end{align} 
where the set $\lbrace c_{a,n} \rbrace$ consists of $N$ real numbers for each value of $a$. Let $ \lbrace  \psi_{a,n}^s 
(k_x, k_y, k_{z_a} )\rbrace $ represent the set of orthonormal wavefunctions associated with $ \lbrace \varepsilon_{a,n}^s \rbrace $.
We also have $ 2 N $ solutions for the momentum along the $z$-axis, for a given set of values for the perpendicular momenta $\lbrace k_x, \,k_y \rbrace $, expressed by $\pm k_{z_a, n}$, where
\begin{align}
k_{z_a, n} = \frac{1} {v_{z_a}} \,
\sqrt{ \left(  \frac{ E - V_a } {c_{a,n} }  \right)^2 
- \left( d_x^{(a)} \right)^2-\left( d_z^{(a)} \right)^2
} \,.
\end{align}
We assume that $E > V_1 $ is the energy of an incident quasiparticle propagating towards the positive $z$-axis.

Using the above notations, we take $\Psi_E^{(1)}$ and $\Psi_E^{(2)}$ to be linear combinations of plane wave solutions propagating along the $z$-axis. Suppose the quasiparticles are incident from the left of the junction with the plane wave momentum $ k_{z_1 , n_0} $. Then we have
\begin{align}
\tilde \Psi_E^{(1)} (k_x, k_y, z) & =  \psi_{a,n_0}^+ (k_x, k_y, k_{z_1, n_0}) \, e^{i\,k_{z_1, n_0} \, z}
+ \sum_{n=1}^{N} r_n \,
{\psi_{a,n}^+  (k_x, k_y, -k_{z_1, n}) \, e^{ - i\,k_{z_1, n} \, z}} \,, \nonumber \\
%%%%%%%%%%%%%%%%%%%%%%%%
\tilde \Psi_E^{(2)} (k_x, k_y, z) & =    \sum_{n=1}^{N} t_n \,
{\psi_{a,n}^+  (k_x, k_y, k_{z_2, n})
\, e^{  i\,k_{z_2 , n} \, z}} \left [ 1 -
\Theta \big ( |{\rm Im} \,k_{z_2 , n} | \big )\right ].
\end{align}
The $2 N$ unknown coefficients $\lbrace r_n,  t_n \rbrace $ can be determined by the $2 N$ equations obtained from Eq.~\eqref{eqbccond}. Note that a $  t_n  $ is defined only for a propagating wave, i.e., when $k_{z_2 , n}$ is real.

The carrier current for each of systems is then along the $z$-axis, and is captured by the probability flux operator $\hat{j}_z= \partial_{k_z} \mathcal{H} (\mathbf  k)
= v_z \, {\mathcal{S}}_z\,.$
%For a $2N$-band system, a wavefunction for a given state can be represented by a column vector $\Psi $ consisting of $2N$-components $ \lbrace \psi_\alpha \rbrace $ with $\alpha \in [1,2N]$. 
The expectation value of the probability flux along the $z$-axis is then given by \cite{Deng2020}
\begin{align}
\langle j_z \rangle
= v_z \,\tilde \Psi_E^\dagger \, {\mathcal{S}}_z \,\tilde \Psi_E \, .
\end{align}
%%%%%%%%
Due to the time-independence of the problem, the continuity of the probability flux reduces to $\partial_z \langle j_z \rangle = 0 $. This condition translates into the statement that
\begin{align}
\label{eqj1}
 & \langle j_z^{(1)} \rangle\Big \vert_{z  <0 }
 =  \langle j_z^{(2)} \rangle \Big \vert_{z > 0}
% \nonumber \\ & 
\Rightarrow
 v_{z_1}  \langle
 \tilde \Psi_E^{(1)} (k_x,k_y, z)| \, {\mathcal{S}}_z  \, | \tilde \Psi_E^{(1)} (k_x,k_y, z) \rangle
   \Big \vert_{z < 0}
 = v_{z_2}  \langle
 \tilde \Psi_E^{(1)} (k_x,k_y, z)| \, {\mathcal{S}}_z  \, | \tilde \Psi_E^{(1)} (k_x,k_y, z) \rangle
  \Big \vert_{z > 0} \,.
%%%%%%%%%%%%%%%%%%%%%
\end{align}
Noting that
\begin{align}
v_{z_a}  
\left[ \psi_{a,n}^s  (k_x, k_y, k_{z_a}) \, e^{i \, k_{z_a} \, z}\right ]^\dagger \, 
{\mathcal{S}}_z  \left [ \psi_{a,n}^s  (k_x, k_y, k_{z_a}) \, e^{i \, k_{z_a} \, z}\right ]
 = \frac{   v_{z_a} ^2    k_{z_a} } 
 { \varepsilon_{n, a}^s  }
\equiv \partial_{k_{z_a}} \varepsilon_{a,n}^s \,,
\end{align} 
%%%%%%%%%%%
Eq.~\eqref{eqj1} reduces to
\begin{align}
& \partial_{ k_{z_1}} \varepsilon_{1,n_0}^+ \Big \vert_{k_{z_1} = k_{z_1, n_0}}
- \sum_n |r_n|^2 \, \partial_{ k_{z_1}} \varepsilon_{1,n}^+ \Big \vert_{k_{z_1} = k_{z_1, n}}
=
\sum_n |t_n|^2 \, \partial_{ k_{z_2}} \varepsilon_{2,n}^+ \Big \vert_{k_{z_2} = k_{z_2, n}}
\nonumber \\
%%%%%%%%%%%%%%%
\Rightarrow & \quad
1- {\mathcal R} = {\mathcal T} \,,
\end{align}
where
\begin{align}
\label{eqtT}
& {\mathcal R} = \sum_n \left ( 
\frac{\partial_{ k_{z_1}} \varepsilon_{1,n}^+ \Big \vert_{k_{z_1} = k_{z_1, n}}}
{\partial_{ k_{z_1}} \varepsilon_{1,n_0}^+ \Big \vert_{k_{z_1} = k_{z_1, n_0}}} \right)
 |r_n|^2
%%%%%%%%%%%%%%
\text{ and }  {\mathcal T} = \sum_n \left(
\frac{ \partial_{ k_{z_2}} \varepsilon_{2,n}^+ \Big \vert_{k_{z_2} = k_{z_2, n}}}
{\partial_{ k_{z_1}} \varepsilon_{1,n_0}^+ \Big \vert_{k_{z_1} = k_{z_1, n_0}}}
\right) | t_n|^2  
\end{align}
are the reflection and transmission probabilities, respectively. This is how the continuity of the probability flux comes into play.

%%%%%%%%%%%%%%%%%
\bibliography{biblio_jn}
\end{document}